%% file: main.tex
\documentclass[rmp,onecolumn,preprintnumbers,amsmath,amssymb,groupedaddress,superscriptaddress]
{revtex4}

\usepackage{graphicx}    
\usepackage{psfrag}                        
\usepackage{amsmath,amssymb,mathrsfs}      
\usepackage{subfigure}                     
\usepackage{stmaryrd}                      
\usepackage{bbm}                           

\newcommand{\nn}{\nonumber}
\newcommand{\bra}{\langle}
\newcommand{\ve}{\vert}
\newcommand{\ket}{\rangle}

\begin{document}

\author{M.~Hartmann, G.~Mahler and O.~Hess}
\title{Fundamentals of Nano-Thermodynamics\\
{\small On the minimal length scales, where temperature exists}}

\input{abst}

\maketitle

\tableofcontents

\input{introduction}

\input{general}

\input{spin}

\input{harmonic}

\input{real}

\input{measure}

\input{summary}

\appendix
\input{appspin}

\input{appharm}

%
%
\input{biblio}

\end{document}

%% file: abst.tex
%
%
\begin{abstract}
Recent progress in the synthesis and processing of nano-structured materials and systems calls for
an improved understanding of thermal properties on small length scales. In this context, the question
whether thermodynamics and, in particular, the concept of temperature can apply on the nanoscale is
of central interest. Here we consider a quantum system consisting of a regular chain of elementary
subsystems with nearest neighbour interactions and assume that the total system is in a canonical state
with temperature $T$. We analyse, under what condition the state factors into a product of canonical
density matrices with respect to groups of $n$ subsystems each, and when these groups have the same
temperature $T$. In quantum systems the minimal group size $n_{\textrm{min}}$ depends on the temperature $T$,
contrary to the classical case. As examples, we apply our analysis to a harmonic chain and different types
of Ising spin chains. For the harmonic chain, which successfully describes thermal properties of insulating
solids, our approach gives a first quantitative estimate of the minimal length scale on which temperature
can exist: This length scale is found to be constant for temperatures above the Debye temperature and
proportional to $T^{-3}$ below. We finally apply the harmonic chain model to various materials of relevance
for technical applications and discuss the results. These show that, indeed, high temperatures can exist
quite locally, while low temperatures exist on larger scales only. This has striking consequences:
In quasi 1-dimensional systems, like Carbon-Nanotubes, room temperatures (300 Kelvin) exist on length scales
of 1 $\mu$m, while very low temperatures (10 Kelvin) can only exist on scales larger than 1 mm.
\end{abstract}

%% file: introduction.tex
%
%
\section{Introduction}

Thermodynamics is among the most successfully and extensively applied theoretical concepts in
physics. Notwithstanding, the various limits of its applicability are not fully understood
\cite{GemmerOtte2001,Allahverdyan2000}.

Of particular interest is its microscopic limit.
Down to which length scales can its standard concepts meaningfully be defined and employed?

Besides its general importance, this question has become increasingly relevant recently since
amazing progress in the synthesis and processing of materials with structures on
nanometer length scales has created a demand for better understanding of thermal properties of
nanoscale devices, individual nanostructures and nanostructured materials
\cite{Cahill2003,Williams1986,Varesi1998,Schwab2000}.
Experimental techniques have improved to such an extent that the measurement of thermodynamic
quantities like temperature with a spatial resolution on the nanometer scale seems within reach
\cite{Gao2002,Pothier1997,Aumentado2001}. In particular, the study of thermal transport in nano scale
devices has made greate advances from an experimental point of view,
while the definition of local temperatures
and thus of temperature profiles remains unclear \cite{Cahill2003}.
Carbon nanotubes play an important role in nano technology and nano scale thermal transport.
According to a recent proposal \cite{Gao2002}
a carbon nanotube filled with Gallium  could be used as a thermometer, which has a spatial extension of only
$10$nm.

Physical properties of smaller and smaller system sizes have been the subject of several experimental
studies in recent years. Some examples are: Investigations of the size dependence of the melting
point of atomic clusters \cite{Schmidt1998}, the size dependence of the surface ferromagnetism of Pd fine
particles \cite{Shinohara2003}, existence of large critical currents of superconductors at nanosizes and
an exceptional surface energy of Ag nanoparticles \cite{Nanda2003}.
Also the system size needed to allow for thermalisation has been investigated. Electrons traveling
through a quantum wire only become thermally distributed if the wire has a certain minimum length
\cite{Pothier1997}.

To provide a basis for the interpretation of present day and future experiments in nanoscale physics
and technology and to obtain a better understanding of the limits of thermodynamics,
it is thus indispensable to clarify the applicability of thermodynamical concepts
on small length scales starting from the most fundamental theory at hand, i. e. quantum mechanics.
In this context, one question appears to be particularly important and interesting:
Can temperature be meaningfully defined on nanometer length scales?

The existence of thermodynamical quantities, i. e. the existence of the thermodynamical limit strongly
depends on the correlations between the considered parts of a system.
For short range interactions, the effective interaction between a region in space and its environment
becomes less relevant compared to its internal energy with increasing region size.
The volume of the region grows proportional to its diameter cubed, while its surface grows
proportional to its diameter squared. 
This scaling behavior is used to show that correlations between a region and its environment
become neglegible in the limit of infinite region size and that therefore the thermodynamical limit exists
\cite{Fisher1964,Ruelle1969,Lebowitz1969}.

To explore the minimal region size needed for the application of thermodynamical concepts, situations far
away from the thermodynamical limit should be analysed. On the other hand, effective correlations between
the considered parts need to be small enough \cite{Hartmann2003a}.
The scaling of interactions between parts of a system compared to the energy contained in the parts
themselves thus sets a minimal length scale on which correlations are
still small enough to permit the definition of local temperatures.
It is the aim of this chapter to study this connection quantitatively.
Typical setups where the minimal length scale, on which temperature exists, becomes relevant are measurements
of a temperature profile with very high resolution. 

In recent years, some attempts to generalise the theory of thermodynamics such that it applies to small
systems, have been made \cite{Hill2001a,Hill2001b,Hill2001c,Rajagopal2004}. These attempts take into account that the thermodynamic limit does not apply to such small systems. On the other hand, they fail to
quantitatively analyse the role of the interactions between adjacent systems.

To analyse the question whether temperature exists locally,
one first has to define when temperature is said to exist.
We adopt here the convention, that a local temperature exists if the considered part of the system is in a
canonical state. The motivation of this convention is twofold:

Following the standard arguments of
statistical mechanics, the canonical state is the maximum entropy state of an interacting system and hence
its equilibrium state. The entropy of the system $S$ is thus a function of its internal energy $U$ and
temperature can be defined via $1 / T \equiv \partial S / \partial U$.

A more practical motivation is the following: Temperature is measured via expectation values of observables,
their second moments etc. The canonical distribution is an exponentially decaying
function of energy characterised by one
single parameter, temperature. This implies that there is a one to one mapping between temperature and
the expectation values of observables, by which temperature is measured. Temperature measurements
via different observables thus yield the same result, contrary to distributions with several
parameters. In large systems with a modular structure, the density of states is a strongly growing
function of energy \cite{Tolman1967}. If the distribution were not exponentially decaying,
the product of the denstity of states times the distribution would not have a pronounced peak
and thus physical quantities like energy would not have ``sharp'' values.

To determine when temperature exists locally, we thus have to analyse when an equilibrium state exists
locally, that is when parts of a large system are in a thermal equilibrium state.
The entire system need not be in an equilibrium state itself. However, local equilibrium is not believed to
exist if the total system is far from equilibrium \cite{Kreuzer1984}. It is not very well understood, when
local equilibrium exists without a global one and only a few exact results on this topic are known.
In our approach, we thus focus on systems which are even globally in an equilibrium state.
Nevertheless, our results are expected to be more general, provided the global state is close to an
equilibrium one \cite{Kubo1985}.

Based on the above argument and noting that a quantum description becomes imperative at nanoscopic scales,
the following approach appears to be reasonable:
Consider a large quantum system, brought into a thermal state via interaction with its environment,
divide this system into subgroups and analyse for what
subgroup-size the concept of temperature is still applicable.

This approach, when applied to concrete models, yields quantitative estimates of minimal length scales
on which temperature can exist. Hence, we apply our concept to two classes of models:

Spin chains have recently been subject of extensive studies in condensed matter physics and
quantum information theory.
Thus correlations and possible local temperatures in spin chains are of interest, both
from a theoretical and experimental point of view \cite{Wang2002,Kenzelmann2002}.
We thus study spin chains with respect to our present purpose.

Harmonic lattice models are a standard tool for the description of thermal properties of solids.
We therefore apply our theory to a harmonic chain model to obtain estimates that are expected to be relevant
for real materials and might be tested by experiments.

This chapter is organized as follows: In section \ref{general}, we present the general
theoretical approach which derives two conditions on the effective group interactions and the global
temperature. We discuss the relation to correlations and several scenaria that might occur.
In the following two sections we apply the general consideration to two concrete models and derive
estimates for the minimal subgroup size.
Section \ref{isingchain} considers various types of Ising spin chains, which have a finite energy spectrum
and show similar features as fermionic systems (e.g. correlated electrons).
Section \ref{harmonicchain} then deals with a harmonic chain, a model with an infinite energy spectrum.
Harmonic chains or lattices successfully describe thermal properties of insulating solids.
In the following section \ref{real} we therefore discuss the results for some real materiales modeled by
harmonic chains. As a last point, we discuss the measurability of local tempeartures in section
\ref {measure}. In the concluding section \ref{conclusion},
we finally compare the results for the different models considered and indicate further interesting topics. 
%

%% file: general.tex
%
\section{General Theory} 
\label{general}

We start by defining the Hamiltonian of our chain in the form \cite{Mahler1998},
\begin{equation}\label{hamil}
H = \sum_{i} H_i + I_{i,i+1}
\end{equation}
where the index $i$ labels the elementary subsystems. $H_i$ is the Hamiltonian of subsystem $i$
and $I_{i,i+1}$ the interaction between subsystem $i$ and $i+1$.
We assume periodic boundary conditions.

%
%
%
\begin{figure}
\centering
\psfrag{n}{$n$}
\includegraphics[width=8cm]{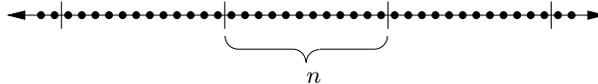}
\caption{Groups of $n$ adjoining subsystems are formed.}
\label{chain}
\end{figure}
We now form $N_G$ groups of $n$ subsystems each as illustrated in figure \ref{chain},
(index $i \rightarrow (\mu-1) n + j; \: \mu = 1, \dots, N_G; \: j = 1, \dots, n$)
and split this Hamiltonian into two parts,
\begin{equation}
\label{hsep}
H = H_0 + I,
\end{equation}
where $H_0$ is the sum of the Hamiltonians of the isolated groups,
\begin{eqnarray}\label{isogroups}
H_0 & = & \sum_{\mu=1}^{N_G} \left( \mathcal{H}_{\mu} - I_{\mu n,\mu n + 1} \right) \enspace \enspace
\textrm{with} \nn \\
\mathcal{H}_{\mu} & = & \sum_{j=1}^n H_{n (\mu - 1) + j} + I_{n (\mu-1) + j,\, n (\mu-1) + j + 1} 
\end{eqnarray}
and $I$ contains the interaction terms of each group with its neighbour group,
\begin{equation}
I = \sum_{\mu=1}^{N_G} I_{\mu n,\mu n + 1}.
\end{equation}
We label the eigenstates of the total Hamiltonian $H$
and their energies with the greek indices $(\varphi, \psi)$ and eigenstates and energies
of the group Hamiltonian  $H_0$ with latin indices $(a, b)$,
\begin{equation}
\label{prodstate}
H \: \ve \varphi \ket = E_{\varphi} \: \ve \varphi \ket \enspace \enspace \textrm{and} \enspace \enspace
H_0 \: \ve a \ket = E_a \: \ve a \ket.
\end{equation}
Here, the states $\ve a \ket$ are products of group eigenstates
\begin{equation}
\ve a \ket = \prod_{\mu = 1}^{N_G} \ve a_{\mu} \ket,
\end{equation}
where
$\left( \mathcal{H}_{\mu} - I_{\mu n, \mu n + 1} \right) \ve a_{\mu} \ket = E_{\mu} \ve a_{\mu} \ket$.
$E_{\mu}$ is the energy of one subgroup only and $E_a = \sum_{\mu=1}^{N_G} E_{\mu}$.

\subsection{Thermal State in the Product Basis}
We assume that the total system is in a thermal state with a density matrix which reads
\begin{equation}
\label{candens}
\bra \varphi \ve \hat \rho \ve \psi \ket = \frac{e^{- \beta E_{\varphi}}}{Z} \: \delta_{\varphi \psi}
\end{equation}
in the eigenbasis of $H$. Here, $Z$ is the partition sum and $\beta = (k_B T)^{-1}$
the inverse temperature with Boltzmann's constant $k_B$ and temperature $T$.
Transforming the density matrix (\ref{candens}) into the eigenbasis of $H_0$ we obtain
\begin{equation}
\label{newrho}
\bra a \ve \hat \rho \ve a \ket =
\int_{E_0}^{E_1} w_a (E) \: \frac{e^{- \beta E}}{Z} \: dE
\end{equation}
for the diagonal elements in the new basis. Here, the sum over all states $\ve a \ket$ has been replaced
by an integral over the energy. $E_0$ is the energy of the ground state and $E_1$ the upper limit of the
spectrum. For systems with an energy spectrum that does not have an upper bound, the limit
$E_1 \rightarrow \infty$ should be taken. The density of conditional probabilities $w_a (E)$ is given by
\begin{equation}
w_a (E) = \frac{1}{\Delta E} \,
\sum_{\{ \ve \varphi \ket: E \le E_{\varphi} < E + \Delta E \}}
\ve \bra a \ve \varphi \ket \ve^2
\end{equation}
where $\Delta E$ is small and the sum runs over all states $\ve \varphi \ket$ with eigenvalues
$E_{\varphi}$ in the interval $[E,E + \Delta E)$.

To compute the integral of equation (\ref{newrho}) we need to know the distribution of the
conditional probabilities $w_a (E)$.
Following the proof of the central limit theorem for mixing sequences \cite{Linnik1971} as a guideline,
one can show that the characteristic function of $H$ does not change if a small subset of
the $\mathcal{H}_{\mu}$ is neglected. Truncating $H$ in a suitable way, the
characteristic function of the remainder of $H$ factorises. If the remainder of $H$ then
fulfills the conditions for Lyapunov's version of the central limit theorem \cite{Billingsley1995} and
$w_a (E)$ converges to a Gaussian normal distribution in the limit of infinite
number of groups $N_G$ \cite{Hartmann2003,Hartmann2004b},
\begin{equation}
\label{gaussian_dist}
\lim_{N_G \to \infty} w_a (E) = \frac{1}{\sqrt{2 \pi} \Delta_a}
\exp \left(- \frac{\left(E - E_a - \varepsilon_a \right)^2}{2 \, \Delta_a^2} \right),
\end{equation}
where the quantities $\varepsilon_a$ and $\Delta_a$ are defined by 
\begin{eqnarray}
\varepsilon_a & \equiv & \bra a \ve H \ve a \ket - \bra a \ve H_0 \ve a \ket \\
\Delta_a^2 & \equiv & \bra a \ve H^2 \ve a \ket - \bra a \ve H \ve a \ket^2.
\end{eqnarray}
Note that $\varepsilon_a$ has a classical counterpart while $\Delta_a^2$ is purely quantum mechanical.
It appears because the commutator $[H,H_0]$ is nonzero, and the distribution $w_a(E)$ therefore has nonzero
width. 

The rigorous proof of equation (\ref{gaussian_dist}) is given in \cite{Hartmann2003} and based on
the following two assumptions:
The energy of each group $\mathcal{H}_{\mu}$ as defined in equation (\ref{isogroups}) is bounded, i. e.
\begin{equation}
\label{bounded}
\bra \chi \ve \mathcal{H}_{\mu} \ve \chi \ket \le C
\end{equation}
for all normalised states $\ve \chi \ket$ and some constant $C$, and
\begin{equation}
\label{vacuumfluc}
\bra a \ve H^2 \ve a \ket - \bra a \ve H \ve a \ket^2 \ge N_G \, C'
\end{equation}
for some constant $C' > 0$. 

Equation (\ref{gaussian_dist}) does not only hold for linear chains, but is equally valid for lattice models
of two or three dimensions. 

In scenarios, where the energy spectrum of each elementary subsystem has an upper limit, such as spins,
condition (\ref{bounded}) is met a priori.
For subsystems with an infinite energy spectrum, such as harmonic oscillators,
we restrict our analysis to states where the energy of every group,
including the interactions with its neighbours, is bounded. Thus, our considerations do not apply
to product states $\ve a \ket$, for which all the energy was located in only one group or only a small
number of groups. The number of such states is vanishingly small compared
to the number of all product states.

If conditions (\ref{bounded}) and (\ref{vacuumfluc}) are met, equation (\ref{newrho})
can be computed for $N_G  \gg 1$:
\begin{align}
\label{newrho2}
\bra a \ve \hat \rho \ve a \ket = & \frac{1}{Z} \,
\exp \left(- \beta \overline{E}_a + \frac{\beta^2 \Delta_a^2}{2} \right) \cdot \nn \\
\cdot & \, \frac{1}{2} \left[\textrm{erfc} \left( \frac{E_0 - \overline{E}_a + \beta \Delta_a^2}{\sqrt{2} \, \Delta_a} \right) -
\textrm{erfc} \left( \frac{E_1 - \overline{E}_a + \beta \Delta_a^2}{\sqrt{2} \, \Delta_a} \right) \right]
\end{align}
where $\overline{E}_a = E_{a} + \varepsilon_a$ and $\textrm{erfc} (x)$ is the conjugate Gaussian error function.
The second term only appears if the energy is bounded and the
integration extends from the energy of the ground state $E_0$ to the upper limit of the spectrum $E_1$.

Note that $\overline{E}_a$ is a sum of $N_G$ terms and that $\Delta_a$ fulfills equation (\ref{vacuumfluc}).
The arguments of the conjugate error functions thus grow proportional to $\sqrt{N_G}$ or stronger.
If these arguments divided by $\sqrt{N_G}$ are finite (different from zero),
the asymptotic expansion of the error function \cite{Abramowitz1970} may thus be used for $N_G \gg 1$:
\begin{equation}
\label{asymptotic_errorf}
\textrm{erfc}(x) \approx \left\{
\begin{array}{lcl}
{\displaystyle \frac{\exp \left(- x^2 \right)}{\sqrt{\pi} \, x}}  & \textrm{for} & x \rightarrow \infty \\
{\displaystyle 2 + \frac{\exp \left(- x^2 \right)}{\sqrt{\pi} \, x}} & \textrm{for} & x \rightarrow - \infty 
\end{array}
\right.
\end{equation}
Inserting this approximation into equation (\ref{newrho2}) and using $E_0 < \overline{E}_a < E_1$ shows
that the second conjugate error function, which contains the upper limit of the energy spectrum,
can always be neglected compared to the first, which contains the ground state energy.

The same type of arguments show that the normalisation of the Gaussian in equation (\ref{gaussian_dist})
is correct although the energy range does not extend over the entire real line ($ -\infty, \infty$).

Applying the asymptotic expansion (\ref{asymptotic_errorf}), equation (\ref{newrho2}) can be taken to read
\begin{equation}
\label{newrho_lower}
\bra a \ve \hat \rho \ve a \ket = \frac{1}{Z}
\exp \left[- \beta \left(E_{a} + \varepsilon_a - \frac{\beta \Delta_a^2}{2} \right) \right]
\end{equation}
for
$\left(E_0 - E_{a} - \varepsilon_a + \beta \Delta_a^2 \right) / \left( \sqrt{2 N_G} \, \Delta_a \right) < 0$
and
\begin{equation} 
\label{newrho_greater}
\bra a \ve \hat \rho \ve a \ket =
\frac{ {\displaystyle \exp \left(- \beta E_0 - \frac{(E_a + \varepsilon_a - E_0)^2}{2 \Delta_a^2} \right)}}
{{\displaystyle \sqrt{2 \pi} \: Z \: \frac{E_0 - E_a - \varepsilon_a + \beta \Delta_a^2}{\Delta_a}}},
\end{equation}
for
$\left(E_0 - E_{a} - \varepsilon_a + \beta \Delta_a^2 \right) / \left( \sqrt{2 N_G} \, \Delta_a \right) > 0$.

The off diagonal elements $\bra a \ve \rho \ve b \ket$ vanish for
$\ve E_a - E_b \ve > \Delta_a + \Delta_b$ because the overlap of the two distributions of conditional
probabilities becomes negligible. For $\ve E_a - E_b \ve < \Delta_a + \Delta_b$, the transformation
involves an integral over frequencies and thus these terms are significantly smaller than
the entries on the diagonal.

\subsection{Correlations}
We now analyse when the thermal state is correlated in the product basis.
To this end, one usually studies the off diagonal elements $\bra a \ve \rho \ve b \ket$ of the density
matrix in the product basis. However, since, in our case, the off diagonal elements
can be estimated to be much smaller than the diagonal elements $\bra a \ve \rho \ve a \ket$,
some information about correlations may be obtained from the latter ones.
For an uncorrelated state, they factorise,
\begin{equation}
\label{factorize}
\bra a \ve \rho \ve a \ket = \prod_{\mu = 1}^{N_G} p_{\mu}
\end{equation}
where $p_{\mu} = \textrm{Tr}_{\overline{\mu}} \bra a \ve \rho \ve a \ket$ and $\textrm{Tr}_{\overline{\mu}}$
is the trace over all groups except group number $\mu$. Taking the logarithm of both sides of
equation (\ref{factorize}) one gets
\begin{equation}
\label{log_factorize}
\ln \left( \bra a \ve \rho \ve a \ket \right) = \sum_{\mu = 1}^{N_G} \ln \left( p_{\mu} \right)
\end{equation}
As follows from equations (\ref{newrho_lower}) and (\ref{newrho_greater}) this can only be the case
for $E_a + \varepsilon_a - E_0 > \beta \Delta_a^2$.
Furthermore, the terms in the exponential on the rhs of equation (\ref{newrho_lower})
must be a sum of terms, each of which depends on the state of one group only.
A term depending on the state $\ve a_{\mu} \ket$ of group $\mu$ must not depend on the state
$\ve a_{\mu+1} \ket$ of the neighbouring group.
In the models we consider here, these conditions coincide with the conditions for the existence of local
temperatures because $\varepsilon_a$ and $\Delta_a^2$ are sums of terms which depend on two neighbouring
groups each.

\subsection{Conditions for Local Thermal States}
We now test under what conditions the diagonal elements of the (local) reduced density-matrices are also canonically distributed with some local inverse temperature $\beta_{\text{loc}}^{(\mu)}$ for each subgroup $\mu = 1, \dots, N_G$.
Since the trace of a matrix is invariant under basis transformations, it is sufficient to verify that they show the correct energy dependence.
If we assume periodic boundary conditions, all reduced density-matrices are equal
($\beta_{\text{loc}}^{(\mu)} = \beta_{\text{loc}}$ for all $\mu$) and the products of their diagonal elements are of the form $\bra a \ve \rho \ve a \ket \propto \exp \left(- \beta_{\text{loc}} E_a \right)$.
We thus have to verify that the logarithm of the right hand side of equations (\ref{newrho_lower}) and
(\ref{newrho_greater}) is a linear function of the energy $E_a$,
\begin{equation} \label{log}
\ln \left( \bra a \ve \hat \rho \ve a \ket \right) \approx - \beta_{\textrm{loc}} \, E_a + c,
\end{equation}
where $\beta_{\textrm{loc}}$ and $c$ are constants. Note that equation (\ref{log}) does not imply that the
occupation probability of an eigenstate $\ve \varphi \ket$ with energy $E_{\varphi}$ and a product state with
the same energy $E_a \approx E_{\varphi}$ are equal. Even if $\beta_{\textrm{loc}}$ and $\beta$ are equal with very
good accuracy, but not exactly the same, occupation probabilities may differ by several orders of magnitude,
provided that the energy range is large enough.

We exclude negative temperatures ($\beta > 0$).
Equation (\ref{log}) can only be true for
\begin{equation} \label{cond_const}
\frac{E_a + \varepsilon_a  - E_0}{\sqrt{N_G} \, \Delta_a} > \beta \frac{\Delta_a^2}{\sqrt{N_G} \, \Delta_a} ,
\end{equation}
as can be seen from equations (\ref{newrho_lower}) and (\ref{newrho_greater}).
Furthermore $\varepsilon_a$ and $\Delta_a^2$ have to be of the following form:
\begin{equation}
\label{cond_linear_1} 
- \varepsilon_a + \frac{\beta}{2} \, \Delta_a^2 \approx  c_1 E_a + c_2 
\end{equation}
where $c_1$ and $c_2$ are constants.
Note that $\varepsilon_a$ and $\Delta_a^2$ need not be functions of $E_a$ and therefore in general
cannot be expanded in a Taylor series.

To ensure that the density matrix of each subgroup $\mu$ is approximately canonical, one needs to satisfy
(\ref{cond_linear_1}) for each subgroup $\mu$ seperately;
\begin{equation}
\label{cond_linear_2} 
- \frac{\varepsilon_{\mu - 1} + \varepsilon_{\mu}}{2} + \frac{\beta}{4} \,
\left(\Delta_{\mu - 1}^2 + \Delta_{\mu}^2 \right)
+ \frac{\beta}{6} \, \tilde{\Delta}_{\mu}^2 \, \approx \, c_1  \, E_{\mu} + c_2
\end{equation}
where 
\begin{eqnarray}
\varepsilon_{\mu} & = & \bra a \ve I_{\mu n, \mu n + 1} \ve a \ket \enspace \enspace \textrm{with} \enspace \enspace 
\varepsilon_a = \sum_{\mu=1}^{N_G} \varepsilon_{\mu}\\
\Delta_{\mu}^2 & = & \bra a \ve \mathcal{H}_{\mu}^2 \ve a \ket -
\bra a \ve \mathcal{H}_{\mu} \ve a \ket^2 \enspace \enspace \textrm{with} \enspace \enspace 
\Delta_a^2 = \sum_{\mu=1}^{N_G} \Delta_{\mu}^2\\
\tilde{\Delta}_{\mu}^2 & = & \sum_{\nu = \mu-1}^{\mu+1} \bra a \ve \mathcal{H}_{\nu-1} \mathcal{H}_{\nu} +
\mathcal{H}_{\nu} \mathcal{H}_{\nu-1} \ve a \ket -
2 \bra a \ve \mathcal{H}_{\nu-1} \ve a \ket \bra a \ve \mathcal{H}_{\nu} \ve a \ket.
\end{eqnarray}

Temperature becomes intensive, if the constant $c_1$ vanishes,
\begin{equation} \label{intensivity}
\left| c_1 \right| \ll 1 \enspace \enspace \Rightarrow \enspace \enspace \beta_{\textrm{loc}} = \beta.
\end{equation}
If this was not the case, temperature would not be intensive, although it might exist locally.

It is sufficient to satisfy conditions (\ref{cond_const}) and (\ref{cond_linear_2}) for an adequate energy
range $E_{\textrm{min}} \le E_{\mu} \le E_{\textrm{max}}$ only.

%
%
%
\begin{figure}
\centering
\psfrag{eta}{\small \hspace{-0.45cm} \raisebox{0.5cm}{$\eta (E)$}}
\psfrag{pb}{\small \hspace{+0.15cm} \raisebox{0.05cm}{$\bra \varphi \ve \rho \ve \varphi \ket$}}
\psfrag{prod}{\small \hspace{0.7cm} \raisebox{0.1cm}{$\eta (E) \cdot \bra \varphi \ve \rho \ve \varphi \ket$}}
\psfrag{E1}{\small \hspace{-0.4cm} $E_{\textrm{min}}$}
\psfrag{E2}{\small \hspace{-0.25cm} $E_{\textrm{max}}$}
\psfrag{n}{}
\psfrag{c1}{$\: E$}
\includegraphics[width=8cm]{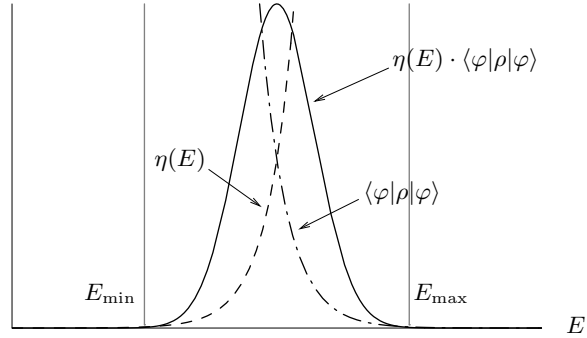}
\caption{The product of the density of states $\eta (E)$ times the occupation probabilities
$\bra \varphi \ve \rho \ve \varphi \ket$ forms a strongly pronounced peak at $E = \overline{E}$.}
\label{peak}
\end{figure}
For large systems with a modular structure, the density of states is
typically a rapidly growing function of energy \cite{Tolman1967}. If the total system is in a
thermal state, occupation probabilities decay exponentially with energy. The product of these two
functions is thus sharply peaked at the expectation value of the energy $\overline{E}$ of the total
system $\overline{E} + E_0 = $Tr$(H \hat \rho)$, with $E_0$ being the ground state energy.
The energy range thus needs to be centered at this peak and large enough.
On the other hand it must not be larger than the range of values $E_{\mu}$ can take on.
Therefore a pertinent and ``save'' choice for $E_{\textrm{min}}$ and $E_{\textrm{max}}$ is
\begin{equation} \label{e_range}
\begin{array}{rcl}
E_{\textrm{min}} & = & \textrm{max}
\left( \left[E_{\mu}\right]_{\textrm{min}} \, , \,
\frac{1}{\alpha} \frac{\overline{E}}{N_G} + \frac{E_0}{N_G} \right)\\
E_{\textrm{max}} & = & \textrm{min}
\left( \left[E_{\mu}\right]_{\textrm{max}} \, , \, \alpha
\frac{\overline{E}}{N_G} + \frac{E_0}{N_G} \right)
\end{array}
\end{equation}
where $\alpha \gg 1$ and $\overline{E}$ will in general depend on the global temperature.
In equation (\ref{e_range}), $\left[E_{\mu}\right]_{\textrm{min}}$ and $\left[E_{\mu}\right]_{\textrm{max}}$ denote
the minimal and maximal values $E_{\mu}$ can take on.

\begin{figure}
\centering
\psfrag{3}{\small \hspace{+0.2cm} $ $}
\psfrag{4}{\small \hspace{+0.2cm} $ $}
\psfrag{5}{\small \raisebox{-0.1cm}{
$ $}}
\psfrag{6}{\small \raisebox{-0.1cm}{
$ $}}
\psfrag{7}{\small \raisebox{-0.1cm}{
$ $}}
\psfrag{8}{\small \raisebox{-0.1cm}{
$ $}}
\psfrag{E1}{\small \hspace{-0.1cm} $E_1$}
\psfrag{E2}{\small \hspace{-0.2cm} $E_2$}
\psfrag{n}{\hspace{-0.3cm} \raisebox{0.1cm}{$\ln \left( \bra a \ve \rho \ve a \ket \right)$}}
\psfrag{c1}{$\: E$}
\includegraphics[width=8cm]{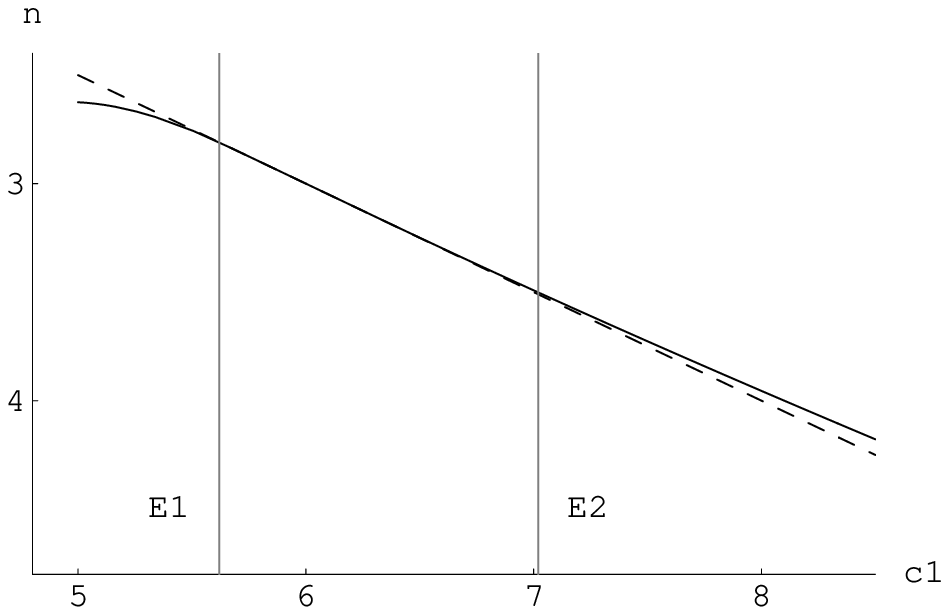}
\caption{$\ln \left( \bra a \ve \rho \ve a \ket \right)$ for $\rho$ as in equation (\ref{newrho2}) (solid line)
and a canonical density matrix $\rho$ (dashed line) for a harmonic chain.}
\label{visual}
\end{figure}

Figure \ref{visual} shows the logarithm of equation (\ref{newrho2}) and the logarithm of a
canonical distribution with the same $\beta$ for a harmonic chain. The actual
density matrix is more mixed than the canonical one.
In the interval between the two vertical lines, both criteria
(\ref{cond_const}) and (\ref{cond_linear_2}) are satisfied.
For $E < E_1$ (\ref{cond_const}) is violated and (\ref{cond_linear_2}) for $E > E_2$. To allow
for a description by means of canonical density matrices, the
group size needs to be chosen such that $E_1 < E_{\textrm{min}}$ and $E_2 > E_{\textrm{max}}$.
%
%
%

For a model obeying equations (\ref{bounded}) and (\ref{vacuumfluc}), the two conditions
(\ref{cond_const}) and (\ref{cond_linear_2}), which constitute the general result of this article,
must both be satisfied. These fundamental criteria will now be applied to some concrete examples.
%


%% file: spin.tex
%
\section{Ising Spin Chain in a Transverse Field\label{isingchain}} 

In this section we consider an Ising spin chain in a transverse field. For this model the Hamiltonian
reads
\begin{eqnarray} \label{ising_ham}
H_i & = & - B \, \sigma_i^z \nn \\
I_{i, i+1} & = & - \frac{J_x}{2} \, \sigma_i^x \otimes \sigma_{i+1}^x -
\frac{J_y}{2} \, \sigma_i^y \otimes \sigma_{i+1}^y
\end{eqnarray}
where $\sigma_i^x, \sigma_i^y$ and $\sigma_i^z$ are the Pauli matrices. $B$ is the magnetic field
and $J_x$ and $J_y$ are two coupling parameters. We will always assume $B > 0$.

The entire chain with periodic boundary conditions may be diagonalized via succesive Jordan-Wigner,
Fourier and Bogoliubov transformations (see appendix \ref{diagon_ising_chain}). The relevant energy scale
is introduced via the thermal expectation value (without the ground state energy)
\begin{equation}
\label{ising_e_quer}
\overline{E} = \frac{n N_G}{2 \pi} \int_{-\pi}^{\pi} dk \,
\frac{\omega_k}{\exp \left( \beta \, \omega_k \right) + 1}
\end{equation}
where $\omega_k$ is given in equation (\ref{ising_frequ}).
The ground state energy $E_0$ is given by
\begin{equation}
\label{ising_e_0}
E_0 = - \frac{n N_G}{2 \pi}
\int_{-\pi}^{\pi} dk \, \frac{\omega_k}{2}.
\end{equation}
Since $N_G \gg 1$, the sums over all modes have been replaced by integrals.

If one partitions the chain into $N_G$ groups of $n$ subsystems each, the groups may also be diagonalized
via a Jordan-Wigner and a Fourier transformation (see appendix \ref{diagon_ising_chain}).
Using the abbreviations  
\begin{equation}
\label{abbreviations}
K = \frac{J_x + J_y}{2 \, B} \enspace \enspace \textrm{and} \enspace \enspace L = \frac{J_x - J_y}{2 \, B},
\end{equation}
the energy $E_a$ reads
\begin{equation}
\label{ising_e_a}
E_a = 2 B \, \sum_{\mu = 1}^{N_G} \sum_{k} \left[ 1 - K \cos(k) \right]
\left(n_k^a(\mu) -\frac{1}{2} \right),
\end{equation}
where $k = \pi l / (n+1)$ ($l = 1, 2, \dots, n$) and $n_k^a(\mu)$ is the fermionic occupation
number of mode $k$ of group $\mu$ in the state $\ve a \ket$. It can take on the values $0$ and $1$.

For the Ising model at hand one has $\varepsilon_a = 0$ for all states
$\ve a \ket$, while the squared variance $\Delta_a^2$ reads
\begin{equation}
\Delta_a^2 = \sum_{\mu=1}^{N_G} \Delta_{\mu}^2,
\end{equation}
with
\begin{eqnarray}
\label{ising_delta_a}
\Delta_{\mu}^2 & = & B^2 \, \left( \frac{K^2}{2} + \frac{L^2}{2} - \right.\nn \\
& - & 2 \left( K^2 - L^2 \right)
\left[\frac{2}{n+1} \sum_{k} \sin^2(k) \, \left( n_k^a(\mu) - \frac{1}{2} \right) \right] \cdot \\
& \enspace & \hspace{1.7cm} \cdot \left.
\left[\frac{2}{n+1} \sum_{p} \sin^2(p) \, \left( n_p^a(\mu+1) - \frac{1}{2} \right) \right] 
\right) \nn
\end{eqnarray}
where the $n_k^a(\mu)$ are the same fermionic occupation numbers as in equation (\ref{ising_e_a}).

The conditions for the central limit theorem are met for the Ising chain apart from two
exceptions. Condition (\ref{bounded}) is always fulfilled as the Hamiltonian of a single spin has
finite dimension. As follows from equation (\ref{ising_delta_a}), condition (\ref{vacuumfluc})
is satisfied except for one single state in the case where $J_x = J_y$ ($L = 0$) and $J_x = - J_y$ ($K = 0$)
respectively.
Therefore, for $N_G \gg 1 $ the fraction of states where our theory does not apply is neglegible.

We now turn to analyse conditions (\ref{cond_const}) and (\ref{cond_linear_2}).
Since the spectrum of the Ising chain is limited, there is no approximation analog to the Debye
approximation for the harmonic chain and $\Delta_{\mu}^2$
cannot be expressed in terms of $E_{\mu-1}$ and $E_{\mu}$. 
We therefore approximate (\ref{cond_const}) and (\ref{cond_linear_2}) with simpler expressions.

Let us first analyse condition (\ref{cond_const}). Since it cannot be checked for
every state $\ve a \ket$ we use the stronger condition
\begin{equation}
\label{cond_const_ising}
E_{\mu} - \frac{E_0}{N_G} > \beta \, \left[\Delta_{\mu}^2 \right]_{\textrm{max}},
\end{equation}
instead. This imlies that (\ref{cond_const}) holds for all states $\ve a \ket$.
We require that (\ref{cond_const_ising}) is true for all states with energies in the range (\ref{e_range}).
It is hardest to satisfy for $E_{\mu} = E_{\textrm{min}}$, we thus get the condition on $n$:
\begin{equation}
\label{cond_const_ising2}
n > \beta \, \frac{\left[\Delta_{\mu}^2 \right]_{\textrm{max}}}{e_{\textrm{min}} - e_0},
\end{equation}
where $e_{\textrm{min}} = E_{\textrm{min}} / n$ and $e_0 = E_0 / (n N_G)$.

We now turn to analyse condition (\ref{cond_linear_2}).
Equation (\ref{ising_delta_a}) shows that the $\Delta_{\mu}^2$ do not contain terms which are
proportional to $E_{\mu}$. One thus has to determine, when the $\Delta_{\mu}^2$ are approximately constant.
This is the case if
\begin{equation}
\label{min_max_linear}
\beta \, \frac{\left[ \Delta_{\mu}^2 \right]_{\textrm{max}} - \left[ \Delta_{\mu}^2 \right]_{\textrm{min}}}{2} \ll
\left[ E_{\mu} \right]_{\textrm{max}} - \left[ E_{\mu} \right]_{\textrm{min}},
\end{equation}
where $[ x ]_{\textrm{max}}$ and $[ x ]_{\textrm{min}}$ denote the maximal and minimal value $x$ takes on in all states
$\ve a \ket$. As a direct consequence, we get
\begin{equation}
\left| c_1 \right| \ll 1
\end{equation}
which means that temperature is intensive. Defining the quantity $e_{\mu} = E_{\mu} / n$, we can rewrite
(\ref{min_max_linear}) as a condition on $n$,
\begin{equation}
\label{min_max_linear2}
n \ge
\frac{\beta}{2 \, \delta} \, \frac{\left[ \Delta_{\mu}^2 \right]_{\textrm{max}} - \left[ \Delta_{\mu}^2 \right]_{\textrm{min}}}
{\left[ e_{\mu} \right]_{\textrm{max}} - \left[ e_{\mu} \right]_{\textrm{min}}}
\end{equation}
where the accuracy parameter $\delta \ll 1$ is equal to the ratio of the lhs and the rhs of
(\ref{min_max_linear}).

Since equation (\ref{min_max_linear}) does not take into account the energy range (\ref{e_range}),
its application needs some further discussion.

If the occupation number of one mode of a group is changed, say from $n_k^a(\mu) = 0$ to $n_k^a(\mu) = 1$,
the corresponding $\Delta_{\mu}^2$ differ at most by $4 \, B^2\, \left| K^2 - L^2 \right| \, / \, (n+1)$.
On the other hand,
$\left[ \Delta_{\mu}^2 \right]_{\textrm{max}} - \left[ \Delta_{\mu}^2 \right]_{\textrm{min}} = B^2 \, \left| K^2 - L^2 \right|$.
The state with the maximal $\Delta_{\mu}^2$ and the state with the minimal $\Delta_{\mu}^2$ thus differ
in nearly all occupation numbers and therefore their difference in energy is close to
$\left[ E_{\mu} \right]_{\textrm{max}} - \left[ E_{\mu} \right]_{\textrm{min}}$. On the other hand,
states with similar energies $E_{\mu}$ also have a similar $\Delta_{\mu}^2$.
Hence the $\Delta_{\mu}^2$ only change quasi continuously with energy and equation (\ref{min_max_linear})
ensures that the $\Delta_{\mu}^2$ are approximately constant even on only a part of the possible energy range.

To illustrate the scenaria that can occur, we are now going to discuss three special coupling models which
represent extremal cases of the possible couplings in our model \cite{Hartmann2004a}.

%
%
\subsection{A coupling with constant $\Delta_a$: $J_y = 0$}

If one of the coupling parameters vanishes ($J_x = 0$ or $J_y = 0$), $K = L$ and $\Delta_{\mu}^2 = B^2 \, K^2$
is constant.
In this case only criterion (\ref{cond_const}) has to be satisfied,
which then coincides with (\ref{cond_const_ising2}).

\begin{figure}
\centering
\psfrag{-8.1}{\small \raisebox{-0.1cm}{$10^{-8}$}}
\psfrag{-6.1}{\small \raisebox{-0.1cm}{$10^{-6}$}}
\psfrag{-4.1}{\small \raisebox{-0.1cm}{$10^{-4}$}}
\psfrag{-2.1}{\small \raisebox{-0.1cm}{$10^{-2}$}}
\psfrag{2.1}{\small \raisebox{-0.1cm}{$10^{2}$}}
\psfrag{2}{\small \hspace{+0.2cm} $10^{2}$}
\psfrag{4}{\small \hspace{+0.2cm} $10^{4}$}
\psfrag{6}{\small \hspace{+0.2cm} $10^{6}$}
\psfrag{8}{\small \hspace{+0.2cm} $10^{8}$}
\psfrag{n}{\raisebox{0.1cm}{$n_{\textrm{min}}$}}
\psfrag{c1}{$\: T / B$}
\includegraphics[width=8cm]{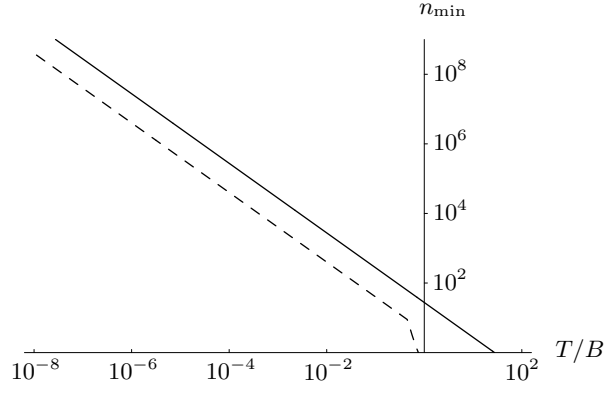}
\caption{Log-log-plot of $n_{\textrm{min}}$ according to eq. (\ref{cond_const_ising2}) for $K = L = 0.1$ (dashed line) and
for $K = L = 10$ (solid line) as a function of $T / B$. $\alpha = 10$ is defined in equation
(\ref{e_range}).}
\label{K=L}
\end{figure}

Inserting expressions (\ref{ising_e_quer}), (\ref{ising_e_0}) and (\ref{ising_delta_a}) with 
$J_x = J$ and $J_y = 0$ into condition (\ref{cond_const_ising2}), one can now
calculate the minimal number of spins per group.
Figure \ref{K=L} shows $n_{\textrm{min}}$ for weak coupling $K = L = 0.1$ and strong coupling
$K = L = 10$ with $\alpha = 10$ as a function of $T / B$.
We choose units where Boltzmann's constant $k_B$ is 1.
%
%
%

Note that, since $\Delta_{\mu} = const$, condition (\ref{cond_const_ising2}) coincides with
criterion (\ref{cond_const}), so that using (\ref{cond_const_ising2}) does not involve any further
approximations.
%
%
%
\begin{figure}
\centering
\psfrag{-1.01}{\small \hspace{-0.07cm} \raisebox{-0.13cm}{$10^{-1}$}}
\psfrag{-0.51}{\small \hspace{-0.12cm} \raisebox{-0.13cm}{$10^{-0.5}$}}
\psfrag{0.01}{\small \hspace{-0.07cm} \raisebox{-0.13cm}{$10^{0}$}}
\psfrag{0.51}{\small \hspace{-0.12cm} \raisebox{-0.13cm}{$10^{0.5}$}}
\psfrag{1.01}{\small \hspace{-0.07cm} \raisebox{-0.13cm}{$10^{1}$}}
\psfrag{1.51}{\small \hspace{-0.12cm} \raisebox{-0.13cm}{$10^{1.5}$}}
\psfrag{0.02}{\small \hspace{0.05cm} \raisebox{-0.05cm}{$0$}}
\psfrag{1.02}{\small \hspace{0.05cm} \raisebox{-0.05cm}{$1$}}
\psfrag{2.02}{\small \hspace{0.05cm} \raisebox{-0.05cm}{$2$}}
\psfrag{3.02}{\small \hspace{0.05cm} \raisebox{-0.05cm}{$3$}}
\psfrag{4.02}{\small \hspace{0.05cm} \raisebox{-0.05cm}{$4$}}
\psfrag{5.02}{\small \hspace{0.05cm} \raisebox{-0.05cm}{$5$}}
\psfrag{2}{\small \hspace{-0.5cm} $10^{2}$}
\psfrag{1}{\small \hspace{-0.5cm} $10^{1}$}
\psfrag{n}{\hspace{-0.8cm} $n_{\textrm{min}}$}
\psfrag{c1}{\hspace{-0.6cm} \raisebox{-0.1cm}{$T / B$}}
\psfrag{c2}{$K$}
\includegraphics[width=8cm]{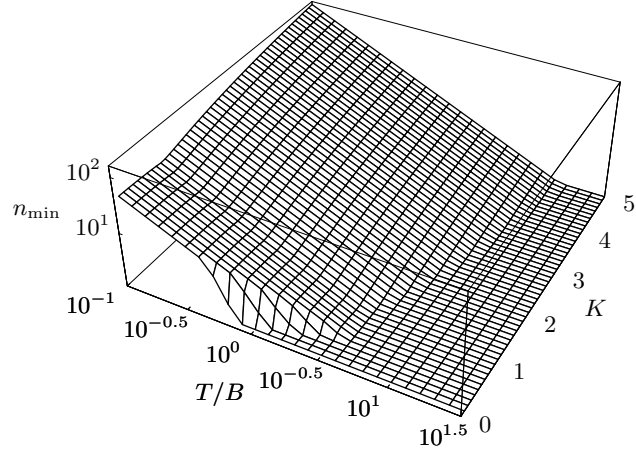}
\caption{$n_{\textrm{min}}$ according to eq. (\ref{cond_const_ising2}) as a function of the coupling
parameter $K$ and $T / B$.
$\alpha = 10$ is defined in equation (\ref{e_range}).}
\label{intvarK=L}
\end{figure}
Figure \ref{intvarK=L} shows $n_{\textrm{min}}$ according to eq. (\ref{cond_const_ising2}) as a function of the coupling
parameter $K$ and $T / B$ for $\alpha = 10$. The range with $K < 1$ represents the weak coupling regime,
since the local level splitting is larger than the coupling strength. As one would expect, $n_{\textrm{min}}$ grows
with increasing coupling $K$. At low temperatures, $n_{\textrm{min}}$ does not smoothly approach $1$ for
$K \rightarrow 0$. This unexpected feature originates from the $N_G \rightarrow \infty$ limit, we consider here.
%
%
\subsection{Fully anisotropic coupling: $J_x = -J_y$}

If both coupling parameters are nonzero, the $\Delta_{\mu}^2$ are not constant. As an example,
we consider here the case of fully anisotropic coupling, where $J_x = - J_y$, i. e. $K = 0$.
Now criteria (\ref{cond_const_ising2}) and (\ref{min_max_linear2}) have to be met.

For $K = 0$, one has $\left[ \Delta_{\mu}^2 \right]_{\textrm{max}} = B^2 \, L^2$,
$\left[ \Delta_{\mu}^2 \right]_{\textrm{min}} = 0$ and
$\left[ e_{\mu} \right]_{\textrm{max}} = - \left[ e_{\mu} \right]_{\textrm{min}} = B$.

\begin{figure}
\centering
\psfrag{-6.1}{\small \raisebox{-0.1cm}{$10^{-6}$}}
\psfrag{-4.1}{\small \raisebox{-0.1cm}{$10^{-4}$}}
\psfrag{-2.1}{\small \raisebox{-0.1cm}{$10^{-2}$}}
\psfrag{2.1}{\small \raisebox{-0.1cm}{$10^{2}$}}
\psfrag{4.1}{\small \raisebox{-0.1cm}{$10^{4}$}}
\psfrag{2}{\small \hspace{+0.2cm} $10^{2}$}
\psfrag{4}{\small \hspace{+0.2cm} $10^{4}$}
\psfrag{6}{\small \hspace{+0.2cm} $10^{6}$}
\psfrag{8}{\small \hspace{+0.2cm} $10^{8}$}
\psfrag{n}{\raisebox{0.1cm}{$n_{\textrm{min}}$}}
\psfrag{c1}{$\: T / B$}
\includegraphics[width=8cm]{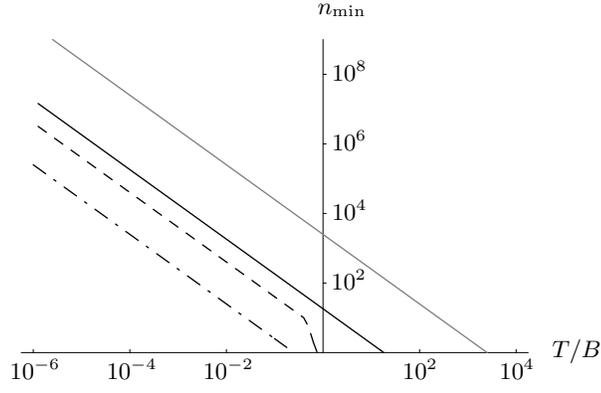}
\caption{Log-log-plot of $n_{\textrm{min}}$ according to eq. (\ref{cond_const_ising2}) for $L = 0.1$ (dashed line) and
for $L = 10$ (solid line) and $n_{\textrm{min}}$ according to eq. (\ref{min_max_linear2}) for $L = 0.1$ (dash - dot line)
and for $L = 10$ (gray line) as a function of $T / B$. $K = 0$, $\alpha = 10$ and $\delta = 0.01$.
$\alpha$ and $\delta$ are defined in equations (\ref{e_range}) and (\ref{min_max_linear2}) respectively.}
\label{K=0}
\end{figure}

We insert these results into (\ref{min_max_linear2}) as well as (\ref{ising_e_quer}) and (\ref{ising_e_0}) 
into (\ref{cond_const_ising2}) and calculate the minimal number of spins per group.
Figure \ref{K=0} shows $n_{\textrm{min}}$ according to criterion (\ref{cond_const_ising2}) and
according to criterion (\ref{min_max_linear2}) seperately, for weak coupling $L = 0.1$ and strong coupling
$L = 10$ with $\alpha = 10$ and $\delta = 0.01$ as a function of $T / B$.
%
%
%

In the present case, all occupation numbers $n_k^a(\mu)$ are zero in the ground state.
In this state, $\Delta_{\mu}^2$ is maximal ($\Delta_{\mu}^2 = B^2 \, L^2$) as can be seen from
(\ref{ising_delta_a}). Therefore criterion (\ref{cond_const_ising2}) is equivalent to
criterion (\ref{cond_const}) for $K = 0$.
%
%
%
\begin{figure}
\centering
\psfrag{-4.01}{\small \hspace{-0.07cm} \raisebox{-0.3cm}{$10^{-4}$}}
\psfrag{-2.01}{\small \hspace{-0.12cm} \raisebox{-0.3cm}{$10^{-2}$}}
\psfrag{0.01}{\small \hspace{-0.07cm} \raisebox{-0.3cm}{$10^{0}$}}
\psfrag{2.01}{\small \hspace{-0.12cm} \raisebox{-0.13cm}{$10^{2}$}}
\psfrag{4.01}{\small \hspace{-0.07cm} \raisebox{-0.13cm}{$10^{4}$}}
\psfrag{0.02}{\small \hspace{0.05cm} \raisebox{-0.05cm}{$0$}}
\psfrag{1.02}{\small \hspace{0.05cm} \raisebox{-0.05cm}{$1$}}
\psfrag{2.02}{\small \hspace{0.05cm} \raisebox{-0.05cm}{$2$}}
\psfrag{3.02}{\small \hspace{0.05cm} \raisebox{-0.05cm}{$3$}}
\psfrag{4.02}{\small \hspace{0.05cm} \raisebox{-0.05cm}{$4$}}
\psfrag{5.02}{\small \hspace{0.05cm} \raisebox{-0.05cm}{$5$}}
\psfrag{2}{\small \hspace{-0.5cm} $10^{2}$}
\psfrag{4}{\small \hspace{-0.5cm} $10^{4}$}
\psfrag{6}{\small \hspace{-0.5cm} $10^{6}$}
\psfrag{n}{\hspace{-0.8cm} $n_{\textrm{min}}$}
\psfrag{c1}{\hspace{-0.6cm} \raisebox{-0.1cm}{$T / B$}}
\psfrag{c2}{$K$}
\includegraphics[width=8cm]{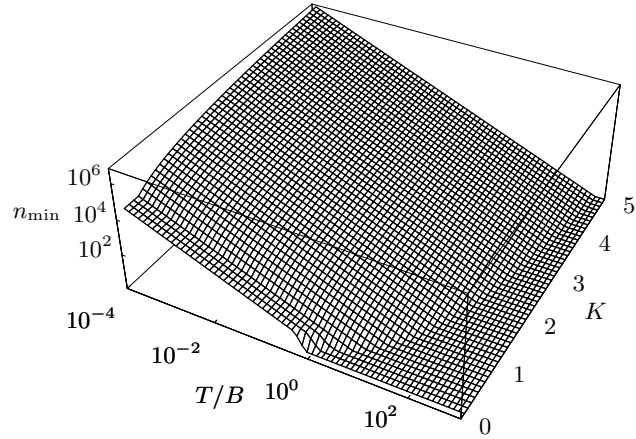}
\caption{$n_{\textrm{min}}$ according to eq. (\ref{cond_const_ising2}) and eq. (\ref{min_max_linear2})
as a function of the coupling parameter $L$ and $T / B$.
$\alpha = 10$ is defined in equation (\ref{e_range}).}
\label{intvarK=0}
\end{figure}

Figure \ref{intvarK=0} shows $n_{\textrm{min}}$ according to eq. (\ref{cond_const_ising2}) as a function of the
coupling parameter $K$ and $T / B$ for $\alpha = 10$. Again, $L < 1$ represents the weak coupling regime,
where the local level splitting is larger than the coupling strength. As in the previous case,
$n_{\textrm{min}}$ is a monotonic increasing function of $L$ but shows a non-smooth transition to the uncoupled ($L = 0$) case.
The latter is again due to $N_G \rightarrow \infty$ limit.
%
%
%
\subsection{Isotropic coupling: $J_x = J_y$}

As a second example of models where both coupling parameters are nonzero, we consider the isotropic coupling
case, where $J_x = J_y$, i. e. $L = 0$.
Again, both criteria (\ref{cond_const_ising2}) and (\ref{min_max_linear2}) have to be met.

The values of $\left[ \Delta_{\mu}^2 \right]_{\textrm{max}}$,
$\left[ \Delta_{\mu}^2 \right]_{\textrm{min}}$, $\left[ e_{\mu} \right]_{\textrm{max}}$ and
$\left[ e_{\mu} \right]_{\textrm{min}}$ are given in equations (\ref{e_min_max_k}), (\ref{e_min_max_K})
and (\ref{delta_min_max}).

For the present model with $L = 0$ and $|K| < 1$ all occupation numbers $n_k^a(\mu)$ are zero in the
ground state and thus $\Delta_{\mu}^2 = 0$. As a consequence, condition (\ref{cond_const_ising2})
cannot be used instead of (\ref{cond_const}). We therefore argue as follows:
In the ground state $E_{\mu} - E_0 / N_G = 0$ as well as $\Delta_{\mu}^2 = 0$ and all
occupation numbers $n_k^a(\mu)$ are zero. If one occupation number is then changed from $0$ to $1$,
$\Delta_{\mu}^2$ changes at most by $4 \, B^2 \, K^2 / (n+1)$ and $E_{\mu}$ changes at least by
$2 \, B \, ( 1 - |K| )$. Therefore (\ref{cond_const}) will hold for all states except the ground state
if
\begin{equation}\label{special_cond_const}
n > 2 \, B \, \beta \, \frac{K^2}{1 - |K|}
\end{equation}

If $|K| > 1$, occupation numbers of modes with $\cos (k) < 1 / |K|$ are zero in the ground state
and occupation numbers of modes with $\cos (k) > 1 / |K|$ are one.
$\Delta_{\mu}^2$ for the ground state then is
$\left[ \Delta_{\mu}^2 \right]_{\textrm{gs}} \approx \left[ \Delta_{\mu}^2 \right]_{\textrm{max}} / 2$ and
(\ref{cond_const_ising2}) is a good approximation of condition (\ref{cond_const}).

\begin{figure}
\centering
\psfrag{-6.1}{\small \raisebox{-0.1cm}{$10^{-6}$}}
\psfrag{-4.1}{\small \raisebox{-0.1cm}{$10^{-4}$}}
\psfrag{-2.1}{\small \raisebox{-0.1cm}{$10^{-2}$}}
\psfrag{2.1}{\small \raisebox{-0.1cm}{$10^{2}$}}
\psfrag{4.1}{\small \raisebox{-0.1cm}{$10^{4}$}}
\psfrag{2}{\small \hspace{+0.2cm} $10^{2}$}
\psfrag{4}{\small \hspace{+0.2cm} $10^{4}$}
\psfrag{6}{\small \hspace{+0.2cm} $10^{6}$}
\psfrag{8}{\small \hspace{+0.2cm} $10^{8}$}
\psfrag{10}{\small \hspace{+0.35cm} $10^{10}$}
\psfrag{12}{\small \hspace{+0.35cm} $10^{12}$}
\psfrag{n}{\raisebox{0.1cm}{$n_{\textrm{min}}$}}
\psfrag{c1}{$\: T / B$}
\includegraphics[width=8cm]{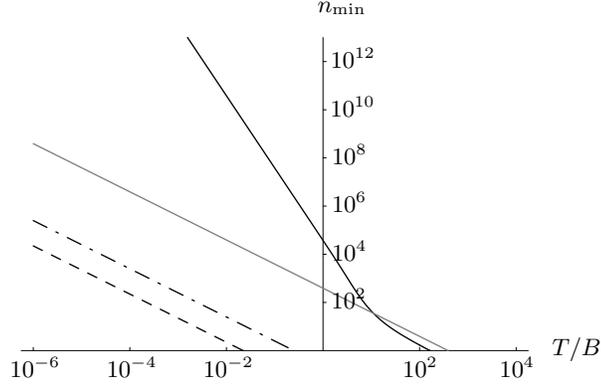}
\caption{Log-log-plot of $n_{\textrm{min}}$ according to eq. (\ref{special_cond_const}) for $K = 0.1$ (dashed line),
according to eq. (\ref{cond_const_ising2}) for $K = 10$ (solid line) and according to eq. (\ref{min_max_linear2}) for $K = 0.1$
(dash - dot line) and for $K = 10$ (gray line) as a function of $T / B$.
$L = 0$, $\alpha = 10$ and $\delta = 0.01$.
$\alpha$ and $\delta$ are defined in equations (\ref{e_range}) and (\ref{min_max_linear2}) respectively.}
\label{L=0}
\end{figure}

Inserting these results into (\ref{min_max_linear2}) as well as (\ref{ising_e_quer}) and (\ref{ising_e_0})
into (\ref{cond_const_ising2}) for $|K| > 1$ and using (\ref{special_cond_const}) for $|K| < 1$,
the minimal number of spins per group can be calculated.
Figure \ref{L=0} shows $n_{\textrm{min}}$ according to criteria (\ref{special_cond_const}) and
(\ref{min_max_linear2}) for weak coupling $K = 0.1$ and
according to criteria (\ref{cond_const_ising2}) and
(\ref{min_max_linear2}) for strong coupling
$K = 10$ with $\alpha = 10$ and $\delta = 0.01$ as a function of $T / B$.
%
%
%
%

%% file: harmonic.tex
%
\section{Harmonic Chain\label{harmonicchain}} 

As a representative for the class of systems with an infinite energy spectrum,
we consider a harmonic chain of $N_G \cdot n$ particles of mass $m$ and
spring constant $\sqrt{m} \, \omega_0$. In this case, the Hamiltonian reads
\begin{eqnarray}
H_i & = & \frac{m}{2} \, p_i^2 + \frac{m}{2} \, \omega_0^2 \, q_{i}^2 \\
I_{i, i+1} & = & - m \, \omega_0^2 \, q_{i} \, q_{i+1},
\end{eqnarray}
where $p_i$ is the momentum of particle $i$ and $q_{i}$ the displacement from its equilibrium
position $i \cdot a_0$ with $a_0$ being the distance between neighbouring particles at equilibrium.
We devide the chain into $N_G$ groups of $n$ particles each and thus get
a partition of the type considered in section~\ref{general}.

The Hamiltonian of one group is diagonalised by a Fourier transform and the definition of creation
and anihilation operators $a_{k}^{\dagger}$ and $a_{k}$ for the Fourier modes (see
appendix \ref{diagon_harmonic_chain}).
\begin{equation}
E_a = \sum_{\mu=1}^{N_G} \sum_{k} \omega_k \left( n_k^a (\mu) + \frac{1}{2} \right),
\end{equation}
where $k = \pi l / (a_0 \, (n+1))$ $(l = 1, 2, \dots, n)$ and the frequencies $\omega_k$ are given by
$\omega^2_{k} = 4 \, \omega_0^2 \, \sin^2(k a / 2)$.
$n_k^a (\mu)$ is the occupation number of mode $k$ of
group $\mu$ in the state $\ve a \ket$. We chose units, where $\hbar = 1$.

To see that this model satisfies the condition (\ref{vacuumfluc}) one needs to
express the group interaction $V(q_{\mu n}, q_{\mu n + 1})$ in terms of
$a_{k}^{\dagger}$ and $a_{k}$, which yields $\tilde{\Delta}_{\mu} = 0$ for all $\mu$
and therefore
\begin{equation}
\Delta_a^2 = \sum_{\mu=1}^{N_G} \Delta_{\mu}^2,
\end{equation}
where $\Delta_{\mu}$, the width of one group interaction, reads
\begin{align}
\label{harmsigma}
\Delta^2_{\mu} = \left( \frac{2}{n+1} \right)^2 &
\left( \sum_{k} \cos^2 \left( \frac{k a_0}{2} \right) \, \omega_{k} \,
\left( n_{k} + \frac{1}{2} \right) \right) \cdot \nn \\
\cdot & \left( \sum_{p} \cos^2 \left( \frac{p a_0}{2} \right) \, \omega_{p} \,
\left( m_{p} + \frac{1}{2} \right) \right).
\end{align}
$\Delta^2_{\mu}$ has a minimum value since all $n_{k} \ge 0$ and all $m_{p} \ge 0$.
In equation (\ref{harmsigma}), $k$ labels the modes of group $\mu$ with occupation numbers $n_k$
and $p$ the modes of group $\mu+1$ with occupation numbers $m_p$.
The width $\Delta^2_a$ thus fulfills condition (\ref{vacuumfluc}).

Since the spectrum of every single oscillator is infinite, condition (\ref{bounded}) can only
be satisfied for states, for which the energy of the system is distributed among a relevant fraction
of the groups, as discussed in section \ref{general}.

The expectation values of the group interactions vanish, $\varepsilon_{\mu} = 0$, while the widths
$\Delta^2_{\mu}$ depend on the occupation numbers $n_{k}$ and therefore on the energies $E_{\mu}$.
We thus apply the conditions (\ref{cond_const}) and (\ref{cond_linear_2}).
To analyse them, we make use of the continuum or Debye approximation \cite{Kittel1983}, requiring
$n \gg 1$, $a_0 \ll l$, where $l = n \, a_0$, and the length of the chain to be finite.
In this case we have $\omega_k = v \, k$ with the constant velocity of sound $v = \omega_0 \, a_0$ and
$\cos (k \, a_0 / 2) \approx 1$. The width of the group interaction (\ref{harmsigma}) thus translates into
\begin{equation}
\label{harmsigma_2}
\Delta^2_{\mu} = \frac{4}{n^2} \, E_{\mu} \, E_{\mu+1}
\end{equation}
where $n+1 \approx n$ has been used. The relevant energy scale is introduced by the thermal expectation
value of the entire chain
\begin{equation}
\label{intenergy}
\overline{E} = N_G n k_B \Theta \left( \frac{T}{\Theta} \right)^2
\int_0^{\Theta / T} \frac{x}{e^x - 1} \, dx,
\end{equation}
and the ground state energy is given by
\begin{equation}
\label{groundenergy}
\overline{E} = N_G n k_B \Theta \left( \frac{T}{\Theta} \right)^2
\int_0^{\Theta / T} \frac{x}{2} \, dx = \frac{N_G n k_B \Theta}{4}
\end{equation}

We first consider criterion (\ref{cond_const}).
For a given $E_a = \sum_{\mu} E_{\mu}$, the squared width $\Delta^2_{\mu}$ is largest if all
$E_{\mu}$ are equal, $x \equiv E_{\mu} \: \forall \mu$. Thus (\ref{cond_const}) is hardest to satisfy for that
case, where it reduces to
\begin{equation}
\label{quadratc_ineq}
x - \frac{E_0}{N_G} - \frac{4 \beta}{n^2} x^2 > 0.
\end{equation}
Equation (\ref{quadratc_ineq}) sets a lower bound on $n$. For $\overline{E} > E_0$, the
bound is strongest for high energies, $x = \alpha (\overline{E} / N_G) + (E_0 / N_G)$, while
for $\overline{E} < E_0$, it is strongest for low energies, $x = (\overline{E} / \alpha  N_G) + (E_0 / N_G)$.
Since condition (\ref{cond_linear_2}) is a stronger criterion than condition (\ref{cond_const})
at $\overline{E} > E_0$, we only consider (\ref{quadratc_ineq}) for low energies, where it reads
\begin{equation}
\label{harmcond2}
n > 4 \, \frac{\Theta}{T} \, \frac{\alpha}{\overline{e}} \, \left(\frac{\overline{e}}{\alpha} + e_0 \right)^2
\end{equation}
with $\overline{e} = \overline{E} / (n N_G k_B \Theta)$ and $e_0 = E_0 / (n N_G k_B \Theta) = 1 / 4$.

To test condition (\ref{cond_linear_2}) we take the derivative with respect to $E_{\mu}$ on both sides,
\begin{equation}
\label{subcondition2}
\frac{\beta}{n^2} \left( E_{\mu-1} + E_{\mu+1} - 2 \, \frac{E_0}{N_G} \right) +
\frac{2 \beta}{n^2} \frac{E_0}{N_G} \approx c_1
\end{equation}
where we have seperated the energy dependent and the constant part on the lhs.
(\ref{subcondition2}) is satisfied if the energy dependent part is much smaller than one, i.e.
\begin{equation}
\label{harmsubcond2}
\frac{\beta}{n^2} \left( E_{\mu-1} + E_{\mu+1} - 2 \frac{E_0}{N_G} \right) \le \delta \ll 1,
\end{equation}
Taking $E_{\mu-1}$ and $E_{\mu+1}$ equal to the upper bound in equation (\ref{e_range}), this yields
\begin{equation}
\label{subcondition3}
n > \frac{2 \alpha}{\delta} \, \frac{\Theta}{T} \, \overline{e},
\end{equation}
where the ``accuracy'' parameter $\delta \ll 1$ quantifies the value of the energy dependent part in
(\ref{subcondition2}).

Since the constant part of the lhs of (\ref{subcondition2}) satisfies
\begin{equation}
\frac{2 \beta}{n^2} \frac{E_0}{N_G} < \frac{\sqrt{\delta}}{\alpha} \,
\left(\frac{1}{\sqrt{2}} - \frac{\sqrt{\delta}}{\alpha} \right) \: \ll 1,
\end{equation}
temperature is intensive.

\begin{figure}
\centering
\psfrag{-4.1}{\small \raisebox{-0.1cm}{$10^{-4}$}}
\psfrag{-3.1}{\small \raisebox{-0.1cm}{$10^{-3}$}}
\psfrag{-2.1}{\small \raisebox{-0.1cm}{$10^{-2}$}}
\psfrag{-1.1}{\small \raisebox{-0.1cm}{$10^{-1}$}}
\psfrag{1.1}{\small \raisebox{-0.1cm}{$10^{1}$}}
\psfrag{2.1}{\small \raisebox{-0.1cm}{$10^{2}$}}
\psfrag{3.1}{\small \raisebox{-0.1cm}{$10^{3}$}}
\psfrag{4.1}{\small \raisebox{-0.1cm}{$10^{4}$}}
\psfrag{1}{}
\psfrag{2}{\small \hspace{+0.2cm} $10^{2}$}
\psfrag{3}{}
\psfrag{4}{\small \hspace{+0.2cm} $10^{4}$}
\psfrag{5}{}
\psfrag{6}{\small \hspace{+0.2cm} $10^{6}$}
\psfrag{7}{}
\psfrag{8}{\small \hspace{+0.2cm} $10^{8}$}
\psfrag{n}{\raisebox{0.1cm}{$n_{min}$}}
\psfrag{c1}{$\: T / \Theta$}
\includegraphics[width=8cm]{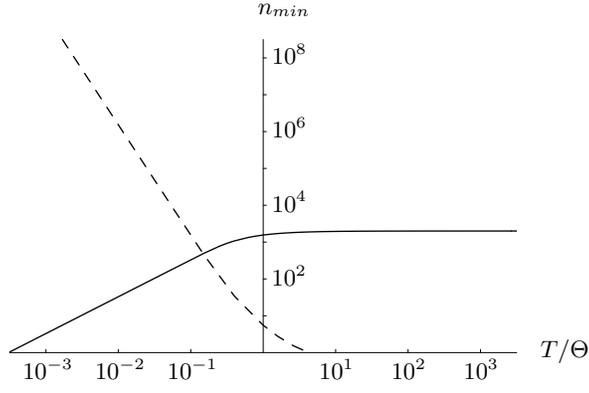}
\caption{Log-log-plot of $n_{min}$ according to eq. (\ref{harmcond2}) (dashed line) and
$n_{min}$ according to eq. (\ref{subcondition3}) (solid line) for $\alpha = 10$ and $\delta = 0.01$
as a function of $T / \Theta$ for a harmonic chain. $\delta$ and $\alpha$ are defined in equations
(\ref{subcondition3}) and (\ref{e_range}), respectively.}
\label{temp}
\end{figure}

Inserting equation (\ref{intenergy}) into equation (\ref{harmcond2}) and (\ref{subcondition3})
one can now calculate the minimal $n$ for given $\delta, \alpha,\Theta$ and $T$.
Figure \ref{temp} shows $n_{min}$ for $\alpha = 10$ and $\delta = 0.01$ given by criterion
(\ref{harmcond2}) and (\ref{subcondition3}) as a function of $T / \Theta$.
%
%
%

\begin{figure}
\centering
\psfrag{-1.1}{\small \hspace{0.1cm} \raisebox{-0.15cm}{$10^{-1}$}}
\psfrag{0.}{\small \hspace{-0.2cm} \raisebox{-0.15cm}{$10^{0}$}}
\psfrag{1.1}{\small \hspace{-0.1cm} \raisebox{-0.15cm}{$10^{1}$}}
\psfrag{2.1}{\small \hspace{-0.1cm} \raisebox{-0.15cm}{$10^{2}$}}
\psfrag{3.1}{\small \hspace{-0.1cm} \raisebox{-0.15cm}{$10^{3}$}}
\psfrag{4.1}{\small \hspace{-0.1cm} \raisebox{-0.15cm}{$10^{4}$}}
\psfrag{5.1}{\small \hspace{-0.1cm} \raisebox{-0.15cm}{$10^{5}$}}
\psfrag{500}{\small \hspace{+0.0cm} $500$}
\psfrag{1000}{\small \hspace{+0.0cm} $1000$}
\psfrag{1500}{\small \hspace{+0.0cm} $1500$}
\psfrag{2000}{\small \hspace{+0.0cm} $2000$}
\psfrag{5}{\small \hspace{-0.5cm} $10^{5}$}
\psfrag{10}{\small \hspace{-0.4cm} $10^{10}$}
\psfrag{n}{\hspace{-0.8cm} \raisebox{0.0cm}{$n_{\textrm{min}}$}}
\psfrag{c1}{\hspace{-1.25cm} \begin{minipage}[t]{2cm} \begin{center}$T$\\ in Kelvin \end{center} \end{minipage}}
\psfrag{c2}{\hspace{+0.2cm} $\: \Theta$ in Kelvin}
\includegraphics[width=8cm]{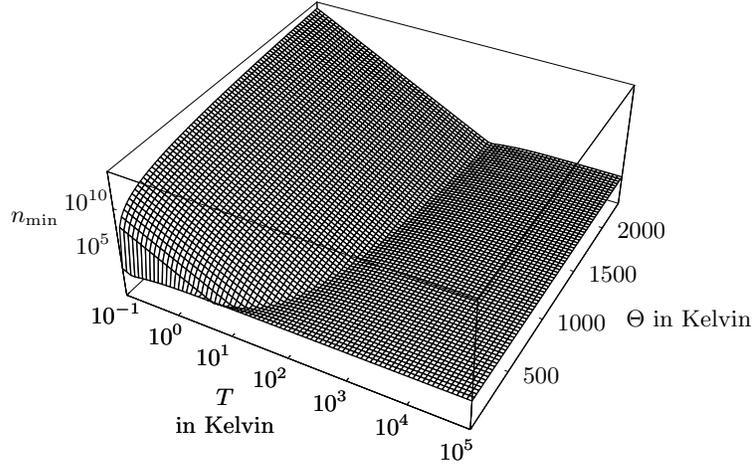}
\caption{Log-log-plot of $n_{\textrm{min}}$ according to eq. (\ref{harmcond2}) and eq. (\ref{subcondition3}) for
$\alpha = 10$ and $\delta = 0.01$ as a function of $T$ and the Debye temperature $\Theta$ for a
harmonic chain. $\delta$ and $\alpha$ are defined in equations
(\ref{subcondition3}) and (\ref{e_range}), respectively.}
\label{matvergl}
\end{figure}

For high (low) temperatures $n_{min}$ can thus be estimated by \cite{Hartmann2003b}
\begin{equation}
n_{min} \approx \left\{
\begin{array}{lcr}
2 \, \alpha / \delta & \textrm{for} & T > \Theta\\
\left( 3 \alpha / 2 \pi^2 \right) \, \left( \Theta / T \right)^3 & \textrm{for} & T < \Theta
\end{array}
\right.
\end{equation}
Figure \ref{matvergl} shows $n_{min}$ for $\alpha = 10$ and $\delta = 0.01$ given by criterion
(\ref{harmcond2}) and (\ref{subcondition3}) as a function of the temperature $T$
and the Debye temperature $\Theta$. As to be expected, $n_{min}$ is a monotonic increasing function of $\Theta$.
%
%
%
%
%


%% file: real.tex
%
\section{Estimates for real materials}
\label{real}

Thermal properties of insulating solids can successfully be described by harmonic lattice models.
Probably the most well known example of such a successfull modeling is the correct prediction
of the temperature dependence of the specific heat based on the Debye theory \cite{Kittel1983}.
We therefor expect our approach for harmonic lattice models to give feasible estimates for real existing
materials.

The minimal length scale on which intensive temperatures exist in insulating solids, is thus given by 
\begin{equation}
\label{length}
l_{\textrm{min}} = n_{\textrm{min}} \, a_0,
\end{equation}
where $a_0$ is the lattice constant, the distance between neighbouring atoms.

Let us consider some materials as examples:

\subsection{Silicon}

\begin{figure}
\centering
\psfrag{-1.1}{\small \raisebox{-0.1cm}{$10^{-1}$}}
\psfrag{1.1}{\small \raisebox{-0.1cm}{$10^{1}$}}
\psfrag{2.1}{\small \raisebox{-0.1cm}{$10^{2}$}}
\psfrag{3.1}{\small \raisebox{-0.1cm}{$10^{3}$}}
\psfrag{4.1}{\small \raisebox{-0.1cm}{$10^{4}$}}
\psfrag{5.1}{\small \raisebox{-0.1cm}{$10^{5}$}}
\psfrag{2}{\small \hspace{-0.6cm} $10^{-8}$}
\psfrag{4}{\small \hspace{-0.6cm} $10^{-6}$}
\psfrag{6}{\small \hspace{-0.6cm} $10^{-4}$}
\psfrag{8}{\small \hspace{-0.6cm} $10^{-2}$}
\psfrag{10}{\small \hspace{-0.0cm} $1$}
\psfrag{n}{\raisebox{0.1cm}{$l_{\textrm{min}}$ in meter}}
\psfrag{c1}{$\: T$ in Kelvin}
\includegraphics[width=8cm]{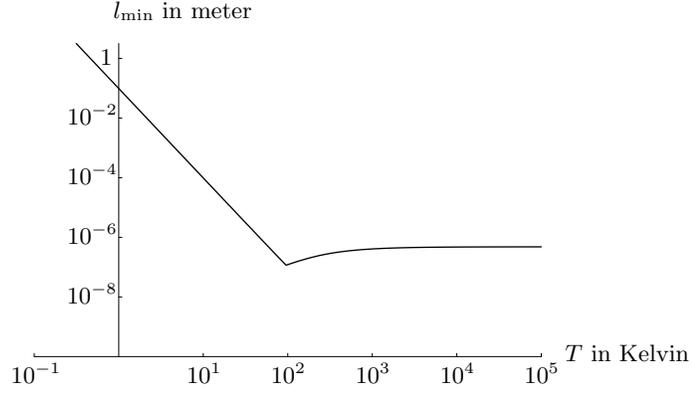}
\caption{$l_{\textrm{min}}$ as a function of temperature $T$ for Silicon.}
\label{temp_Si}
\end{figure}

Silicon is used in many branches of technology. It has a Debye temperature of $\Theta \approx 645 \,$K and
its lattice constant is $a_0 \approx 2.4 \,${\AA}.
Using these parameters,
figure \ref{temp_Si} shows the minimal length-scale on which temperature can exist in a one-dimensional
silicon wire as a function of global temperature.
%
%
%

\subsection{Carbon}

Recently, carbon became increasingly important in the fabrication of nano-structured devices
\cite{Dresselhaus1996,Dresselhaus2001}.
It has a Debye temperature of $\Theta \approx 2230 \,$K and
a lattice constant of $a_0 \approx 1.5 \,${\AA}.

\begin{figure}
\centering
\psfrag{-1.1}{\small \raisebox{-0.1cm}{$10^{-1}$}}
\psfrag{1.1}{\small \raisebox{-0.1cm}{$10^{1}$}}
\psfrag{2.1}{\small \raisebox{-0.1cm}{$10^{2}$}}
\psfrag{3.1}{\small \raisebox{-0.1cm}{$10^{3}$}}
\psfrag{4.1}{\small \raisebox{-0.1cm}{$10^{4}$}}
\psfrag{5.1}{\small \raisebox{-0.1cm}{$10^{5}$}}
\psfrag{2}{\small \hspace{-0.6cm} $10^{-8}$}
\psfrag{4}{\small \hspace{-0.6cm} $10^{-6}$}
\psfrag{6}{\small \hspace{-0.6cm} $10^{-4}$}
\psfrag{8}{\small \hspace{-0.6cm} $10^{-2}$}
\psfrag{10}{\small \hspace{-0.0cm} $1$}
\psfrag{n}{\raisebox{0.1cm}{$l_{\textrm{min}}$ in meter}}
\psfrag{c1}{$\: T$ in Kelvin}
\includegraphics[width=8cm]{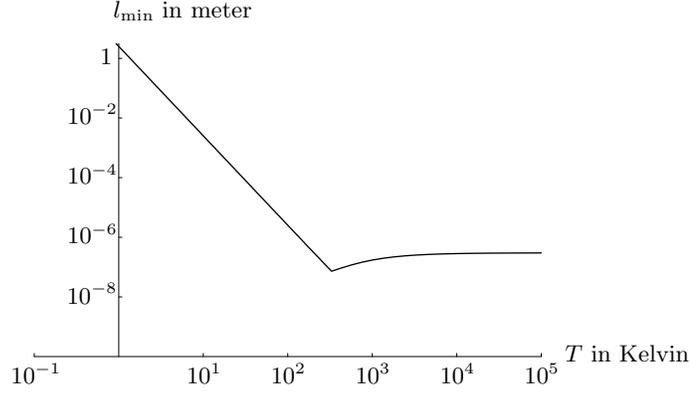}
\caption{$l_{\textrm{min}}$ as a function of temperature $T$ for Carbon.}
\label{temp_C}
\end{figure}

Figure \ref{temp_C} shows the minimal length-scale on which temperature can exist in a one-dimensional
carbon device as a function of global temperature. 

Carbon Nanotubes have diameters of only a few
nanometers. Figure \ref{temp_C} thus provides a good estimate of the maximal accuracy, with which
temperature profiles in such tubes can be meanigfully discussed \cite{Cahill2003}.
%
%
%

\subsection{Gallium, Arsen and Indium}

\begin{figure}
\centering
\psfrag{-1.1}{\small \raisebox{-0.1cm}{$10^{-1}$}}
\psfrag{1.1}{\small \raisebox{-0.1cm}{$10^{1}$}}
\psfrag{2.1}{\small \raisebox{-0.1cm}{$10^{2}$}}
\psfrag{3.1}{\small \raisebox{-0.1cm}{$10^{3}$}}
\psfrag{4.1}{\small \raisebox{-0.1cm}{$10^{4}$}}
\psfrag{5.1}{\small \raisebox{-0.1cm}{$10^{5}$}}
\psfrag{2}{\small \hspace{-0.6cm} $10^{-8}$}
\psfrag{4}{\small \hspace{-0.6cm} $10^{-6}$}
\psfrag{6}{\small \hspace{-0.6cm} $10^{-4}$}
\psfrag{8}{\small \hspace{-0.6cm} $10^{-2}$}
\psfrag{10}{\small \hspace{-0.0cm} $1$}
\psfrag{n}{\raisebox{0.1cm}{$l_{\textrm{min}}$ in meter}}
\psfrag{c1}{$\: T$ in Kelvin}
\includegraphics[width=8cm]{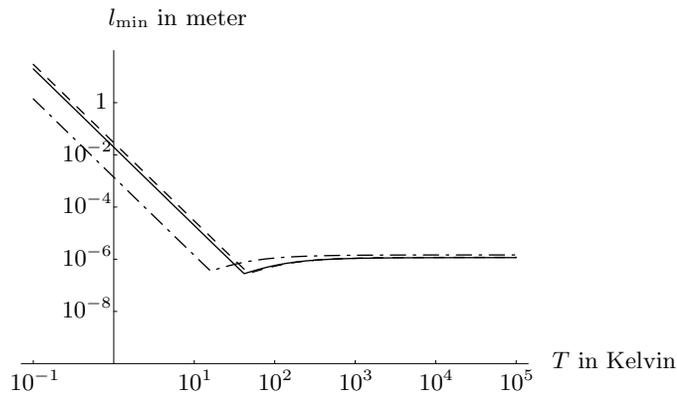}
\caption{$l_{\textrm{min}}$ as a function of temperature $T$ for Gallium (dashed line), Arsen (solid line) and Indium
(dash-dotted line)}
\label{temp_GaAsIn}
\end{figure}

As further representatives of materials which are of central relevance in todays optoelectronics and photonics,
Figure \ref{temp} shows $n_{\textrm{min}}$ for $\alpha = 10$ and $\delta = 0.01$ given by criterion
(\ref{harmcond2}) and (\ref{subcondition3}) as a function of $T / \Theta$.
we consider the semiconductors Gallium ($\Theta \approx 320 \,$K, $a_0 \approx 2.4 \,${\AA}),
Arsen ($\Theta \approx 282 \,$K, $a_0 \approx 3.2 \,${\AA}) and
Indium ($\Theta \approx 108 \,$K, $a_0 \approx 3.3 \,${\AA}).
Figure \ref{temp_GaAsIn} shows $l_{\textrm{min}}$ for those materials.

%
%
%
%


%% file: measure.tex
%
\section{Local temperature measurement \label{measure}}

Having determined the limits on the existence of local temperatures, one is of course interested in
whether and under what conditions temperature can
be measured locally. We thus discuss this question in the present section.

The standard setup for a temperature measurement consists of the system or a part of it,
whose temperature is to be
measured and the thermometer. The latter must be significantly smaller than the system and the interaction
between the system and the thermometer must be weak, so that the exchange of energy between both cannot
significantly alter the proper energy of the system.
The thermometer furthermore should have the property, that there exists a one to one mapping between its
temperature and another quantity, which is easily accessible to the observer, like the height of a meniscus
or the pressure of a gas.
%
%
%
\begin{figure}
\centering
\psfrag{m}{\hspace{-2.2cm}part of the system to be measured}
\psfrag{t}{thermometer}
\includegraphics[width=8cm]{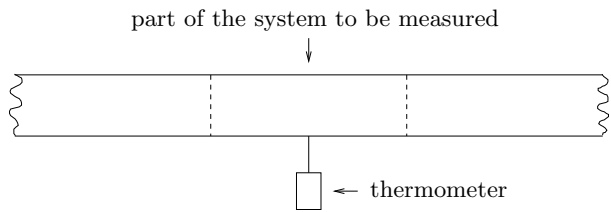}
\caption{Measurement setup}
\label{measuresetup}
\end{figure}

If temperature exists for the system or its part of interest, the respective part is in a thermal 
equilibrium state.
The thermometer then relaxes into an equilibrium state of the same temperature
and thus measures the latter. From this point of view it is unclear whether the thermometer would
measure the local or global equilibrium state.

However, if one views the thermometer as a part of the system, the situation becomes rather obvious.
If there exist an equilibrium state in the part of the system which contains
the thermometer, this part can be described by a canonical density matrix.
Since the interaction between the system (without the thermometer) and the thermometer is very weak this
density matrix factorises
into two canonical ones of the same temperature, one for the system and one for the thermometer.
Therefore, the thermometer measures the temperature on a length scale where the ``reduced'' density matrix is
closest to a canonical one.

For a system in global equilibrium, the thermometer will thus always measure the global temperature.
On the other hand, if equilibrium exists only locally, the thermometer measures the local temperature.

The length scales derived in this chapter are thus only measurable, by standard methods,
in scenarios of global non-equilibrium.
It is therefore important to note that our results, although they are derived for global equilibrium scenarios,
should also apply in more general settings.

Whether local temperature can be measured in global equilibrium cases, e.g. via a measurement
of occupation probabilities of the phonon spectrum, is an interesting subject of future research.


%% file: summary.tex
%
\section{Discussion and Conclusions\label{conclusion}}

In summary, we have analysed the limits of the existence of temperature on the nanoscale.
To this end, we have considered a linear chain of particles interacting with their nearest neighbours.
We have partitioned the chain into identical groups of $n$ adjoining particles each.
Taking the number of such groups to be very large and assuming the total
system to be in a thermal state with temperature $T$ we have found conditions
(equations (\ref{cond_const}) and (\ref{cond_linear_2})), which ensure that each group is approximately
in a thermal state. Furthermore, we have determined when the isolated groups have the same temperature $T$,
that is when temperature is intensive.

The result shows that, in the quantum regime,
these conditions depend on the temperature $T$, contrary to the classical case.
The characteristics of the temperature dependence are determined by the energy dependence of $\Delta_a$.
The low temperature behavior, in particular, is related to the fact that $\Delta_a$ has a nonzero
minimal value. This fact does not only appear in the harmonic chain or spin chains 
but is a general quantum feature of most systems of that topology.
The commutator $[H,H_0]$ is nonzero and the ground state of the total
system is energetically lower than the lowest product state, therefore $\sigma_a$ is nonzero, even
at zero temperature \cite{Wang2002,Jordan2003,Allahverdyan2002,Nieuwenhuizen2002}.

We have then applied the general method to a harmonic chain and several types of Ising spin chains.
For concrete models, the conditions (\ref{cond_const}) and (\ref{cond_linear_2}) determine a 
minimal group size and thus a minimal length scale
on which temperature may be defined according to the temperature concept we adopt.
Grains of size below this length scale are no more in a thermal state. Thus temperature measurements
with a higher resolution should no longer be interpreted in a standard way.

The most striking difference between the spin chains and the harmonic chain is that the energy spectrum
of the spin chains is limited, while it is infinite for the harmonic chain.
For spins at very high global temperatures, the total density matrix is then almost completely mixed, i. e.
proportional to the identity matrix, and thus does not change under basis transformations.
There are thus global temperatures which are high enough, so that local temperatures exist even for
single spins.

For the harmonic chain, this feature does not appear, since the size of the relevant energy range
increases indefinitely with growing global temperature, leading to the constant minimal length
scale in the high energy range.

Our approach should be seen as a first estimate of these minimal length scales. It is thoroughly derived,
nonetheless, the treatement may still be made more precise in some points.
On the one hand, an exact treatement of the off-diagonal elements of the density matrix would give more
justification to the results. On the other hand, the choice of the relevant energy range (\ref{e_range})
could be made more precise, taking into account the shape of the peak (see figure \ref{peak}).

The results we have obtained for the harmonic chain model allow to make predictions for real existing materials.
We have thus considered some materials, which play an important role in fields of current technological
interest.

We have then discussed the measurability of local temperatures in a standard scenario and found that
a thermometer always measures temperature on a length scale, where the ``reduced'' density matrix is closest to
a canonical one, i.e. closest to an equilibrium state. Hence a thermometer measures the global temperature as
long as the total system is in equilibrium, while it measures the local one if equilibrium exist only locally.  

The length scales, calculated in this paper, should also constrain the way one can meaningfully
define temperature profiles in non-equilibrium scenarios \cite{Michel2003}.
Although, our approach only considers cases where even the total system is in an equilibrium state,
the results are expected to be more general.

A typical non-equilibrium scenario would be a heat conduction experiment, where
the two ends of a wire are conected to two heat baths at different temperatures. In this setup, a temperature
gradient would build up in the wire. If the gradient is small, the density matrix of the wire would be somewhat
similar to a canonical one. Applying our approach to this quasi-equilibrium case, would then yield results of
the same order of magnitude as the present treatment of the full equilibrium case.


%% file: appspin.tex
%
\section{Diagonalization of the Ising Chain \label{diagon_ising_chain}}

The Hamiltonian of the Ising chain is diagonalised via Jordan-Wigner transformation which maps it
to a fermionic system.
\begin{eqnarray}
c_i & = & \left( \prod_{j < i} \sigma_j^z \right) \frac{\sigma_i^x + i \sigma_i^y}{2} \nn \\
c_i^{\dagger} & = & \left( \prod_{j < i} \sigma_j^z \right) \frac{\sigma_i^x - i \sigma_i^y}{2}
\end{eqnarray}
The operators $c_i$ and $c_i^{\dagger}$ fulfill fermionic anti-commutation relations
\begin{eqnarray}
\label{fermionic_comm}
\{c_i, c_j \} & = & 0 \nn \\
\{c_i, c_j^{\dagger} \} & = & \delta_{i j}
\end{eqnarray}
and the Hamiltonian reads 
\begin{equation} \label{fermionic_ising_ham}
H = B \left[ \sum_j \left(2 c_j^{\dagger} c_j - 1 \right) -
K \sum_j \left( c_j^{\dagger} c_{j+1} + \textrm{h.c.} \right) -
L \sum_j \left( c_j^{\dagger} c_{j+1}^{\dagger} + \textrm{h.c.} \right) \right]
\end{equation}
with $K = (J_x + J_y) / (2 B)$ and $L = (J_x - J_y) / (2 B)$.
In the case of periodic boundary conditions a boundary term is neglected in equation
(\ref{fermionic_ising_ham}).
For long chains ($n N_G \rightarrow \infty$) this term is supressed by a factor $(n N_G)^{-1}$.
The Hamiltonian now describes Fermions which interact with their nearest neighbours.
As for the bosonic system, a Fourier transformations maps the system to noninteracting fermions.
For the whole chain with periodic boundary conditions
\begin{equation}
\left\{
\begin{array}{c}
c_j^{\dagger} \\ c_j
\end{array} \right\}
= \frac{1}{\sqrt{n N_G}} \sum_k e^{i k j}
\left\{
\begin{array}{c}
d_k^{\dagger} \\ d_k
\end{array} \right\}
\end{equation}
with $k = (2 \pi l) / (n N_G)$ where
$l = 0, \pm 1, \dots, \pm (n N_G - 2) / 2, \, (n N_G) / 2$ for $n N_G$ even,
and
\begin{equation}
\left\{
\begin{array}{c}
c_j^{\dagger} \\ c_j
\end{array} \right\}
= \sqrt{\frac{2}{n + 1}} \sum_k \sin (k j)
\left\{
\begin{array}{c}
d_k^{\dagger} \\ d_k
\end{array} \right\}
\end{equation}
with $k = (\pi l) / (n + 1)$ and ($l = 1, 2, \dots, n$) for one single group.

In the case of periodic boundary conditions, fermion interactions of the form $d_k^{\dagger} d_{-k}^{\dagger}$
and $d_k d_{-k}$ remain.
Therefore, one still has to apply a Bogoliubov transformation to diagonalize the system, i.e.
\begin{eqnarray}
d_k^{\dagger} & = & u_k b_k^{\dagger} - i v_k b_{-k} \nn \\
d_k & = & u_k b_k + i v_k b_{-k}^{\dagger}
\end{eqnarray}
where $u_k = u_{-k}$, $v_k = - v_{-k}$ and $u_k^2 + v_k^2 = 1$.
With the definitions $u_k = \cos(\Theta_k / 2)$ and $v_k = \sin(\Theta_k / 2)$ the interaction
terms disappear for
\begin{equation}
\cos(\Theta_k) =
\frac{1 - K \cos k}{\sqrt{[1 - K \cos k]^2 + [L \sin k]^2}}
\end{equation}

In the case of the finite chain of one group, the  Bogoliubov transformation is not needed, since
the corresponding terms are of the form $d_k^{\dagger} d_k^{\dagger}$ and $d_k d_k$ and vanish by
virtue of equation (\ref{fermionic_comm}).

The Hamiltonians in the diagonal form read
\begin{equation}
H = \sum_k \omega_k \left( b_k^{\dagger} b_k - \frac{1}{2} \right)
\end{equation}
where the frequencies are
\begin{equation} \label{ising_frequ}
\omega_k = 2 B \sqrt{[1 - K \cos k]^2 + [L \sin k]^2}
\end{equation}
with $k = (2 \pi l) / (n N_G)$ for the periodic chain and
\begin{equation}
\omega_k = 2 B \left( 1 - K \cos k \right)
\end{equation}
with $k = (\pi l) / (n + 1)$ for the finite chain.

For the finite chain the occupation number operators may also
be chosen such that $\omega_k$ is always positive. Here, the convention at hand is more convenient,
since the same occupation numbers also appear in the group interaction and thus in $\Delta_{\mu}$.

\subsection{Maxima and minima of $E_{\mu}$ and $\Delta_{\mu}^2$}

The maximal and minimal values of $E_{\mu}$ are given by
\begin{equation}\label{e_min_max_k}
\left\{ \begin{array}{c} \left[E_{\mu} \right]_{\textrm{max}} \\ \left[E_{\mu}\right]_{\textrm{min}} \end{array} \right\} =
\left\{ \begin{array}{c} + \\ - \end{array} \right\} \, n \, B,
\end{equation}
for $| K | < 1$ and by
\begin{equation}\label{e_min_max_K}
\left\{ \begin{array}{c} \left[E_{\mu} \right]_{\textrm{max}} \\ \left[E_{\mu}\right]_{\textrm{min}} \end{array} \right\} =
\left\{ \begin{array}{c} + \\ - \end{array} \right\} \, n \, B \,
\frac{2}{\pi} \left[ \sqrt{K^2 - 1} + \arcsin \left( \frac{1}{| K |} \right) \right],
\end{equation}
for $| K | > 1$, where the sum over all modes $k$ has been approximated with an integral.

The maximal and minimal values of $\Delta_{\mu}^2$ are given by
\begin{equation}\label{delta_min_max}
\left\{ \begin{array}{c} \left[ \Delta_{\mu}^2 \right]_{\textrm{max}}
\\ \left[ \Delta_{\mu}^2 \right]_{\textrm{min}} \end{array} \right\}
= B^2 \left\{ \begin{array}{c} \textrm{max} \left( K^2 , L^2 \right) \\
\textrm{min} \left( K^2 , L^2 \right) \end{array} \right\}.
\end{equation}
%


%% file: appharm.tex
%
\section{Diagonalization of the Harmonic Chain \label{diagon_harmonic_chain}}

The Hamiltonian of a harmonic chain is diagonalised by a Fourier transformation
and the definition of creation and anihilation operators.

For the entire chain with periodic boundary conditions, the Fourier transformation reads
\begin{equation}
\left\{
\begin{array}{c}
q_{j}\\ p_{j}
\end{array} \right\} =
\frac{1}{\sqrt{n N_G}}
\sum_{k} 
\left\{ \begin{array}{c}
u_{k} \exp (i a_0 k j) \\ v_{k} \exp (- i a_0 k j) 
\end{array} \right\}
\end{equation}
with $k = 2 \pi l / (a_0 \, n \, N_G)$ and $(l = 0, \pm 1, \dots,$ \linebreak $\pm (n N_G - 2) / 2,
\, (n N_G) / 2$,
where $n N_G$ has been assumed to be even.

For the diagonalisation of one single group, the Fourier transformation is
\begin{equation}
\left\{
\begin{array}{c}
q_{j}\\ p_{j}
\end{array} \right\} =
\sqrt{\frac{2}{n+1}}
\sum_{k} 
\left\{ \begin{array}{c}
u_{k}\\ v_{k}
\end{array} \right\}
\sin (a_0 k j)
\end{equation}
with
$k = \pi l / (a_0 \, (n+1))$ and $(l = 1, 2, \dots, n)$.

The definition of the creation and anihilation operators is in both cases
\begin{equation}
\left\{
\begin{array}{c}
a_{k}^{\dagger} \\ a_{k}
\end{array} \right\} =
\frac{1}{\sqrt{2 m \omega_{k}}} \:
\left( m \omega_{k} u_{k}
\left\{
\begin{array}{c}
- \\ +
\end{array} \right\}
i v_{k} \right) 
\end{equation}
where the corresponding $u_{k}$ and $v_{k}$ have to be inserted.
The frequencies $\omega_{k}$ are given by $\omega^2_{k} = 4 \omega_0^2 \sin^2(k a_0 / 2)$ in both cases..

The operators $a_{k}^{\dagger}$ and $a_{k}$ satisfy bosonic commutation relations
\begin{eqnarray}
[a_{k}, a_{p} ] & = & 0 \nn \\
\left[\right.a_{k}, a_{p}^{\dagger} \left.\right] & = & \delta_{k p}
\end{eqnarray}
and the diagonalised Hamiltonian reads
\begin{equation}
H = \sum_k \omega_k \left(a_{k}^{\dagger} a_k + \frac{1}{2} \right)
\end{equation}
%


%% file: biblio.tex

\bibliographystyle{plain}
\bibliography{mybib}


%% file: main.bbl
\begin{thebibliography}{10}

\bibitem{Abramowitz1970}
M.~Abramowitz and I.~Stegun.
\newblock {\em Handbook of {M}athematical {F}unctions}.
\newblock Dover Publ., New York, 9th edition, 1970.

\bibitem{Allahverdyan2000}
A.~E Allahverdyan and Th.~M Nieuwenhuizen.
\newblock Extraction of {W}ork from a {S}ingle {T}hermal {B}ath in the
  {Q}uantum {R}egime.
\newblock {\em Phys. Rev. Lett.}, 85:1799--1802, 2000.

\bibitem{Allahverdyan2002}
A.~E. Allahverdyan and Th.~M. Nieuwenhuizen.
\newblock Testing the {V}iolation of the {C}lausius {I}nequality in {N}anoscale
  {E}lectric {C}ircuits.
\newblock {\em Phys. Rev. B}, 66:115309, 2002.

\bibitem{Aumentado2001}
J.~Aumentado, J.~Eom, V.~Chandrasekhar, P.M. Baldo, and L.E. Rehn.
\newblock Proximity effect thermometer for local electron temperature
  measurements on mesoscopic samples.
\newblock {\em Appl. Phys. Lett.}, 75:3554, 1999.

\bibitem{Billingsley1995}
P.~Billingsley.
\newblock {\em Probability and {M}easure}.
\newblock John Wiley \& Sons, New York, 3rd edition, 1995.

\bibitem{Cahill2003}
D.~Cahill, W.~Ford, K.~Goodson, G.~Mahan, A.~Majumdar, H.~Maris, R.~Merlin, and
  S.~Phillpot.
\newblock Nanoscale thermal transport.
\newblock {\em J. Appl. Phys.}, 93:793, 2003.

\bibitem{Dresselhaus2001}
M.S. Dresselhaus, editor.
\newblock {\em Carbon {N}anotubes}, volume~80 of {\em Topics in Applied
  Physics}.
\newblock Springer, Berlin, 2001.

\bibitem{Dresselhaus1996}
M.S. Dresselhaus, G.~Dresselhaus, and P.C. Eklund.
\newblock {\em Science of {F}ullerenes and {C}arbon {N}anotubes}.
\newblock Academic Press, San Diego, 1996.

\bibitem{Fisher1964}
M.E. Fisher.
\newblock The {F}ree {E}nergy of a {M}acroscopic {S}ystem.
\newblock {\em Arch. Ratl. Mech. Anal.}, 17:377, 1964.

\bibitem{Gao2002}
Y.~Gao and Y.~Bando.
\newblock Carbon nanothermometer containing gallium.
\newblock {\em Nature}, 415:599, 2002.

\bibitem{GemmerOtte2001}
J.~Gemmer, A.~Otte, and G.~Mahler.
\newblock Quantum approach to a derivation of the second law of thermodynamics.
\newblock {\em Phys.\ Rev.\ Lett.}, 86:1927--1930, 2001.

\bibitem{Hartmann2003a}
M.~Hartmann, J.~Gemmer, G.~Mahler, and O.~Hess.
\newblock Scaling behavior of interactions in a modular quantum system and the
  existence of local temperature.
\newblock {\em Euro. Phys. Lett.}, 65:613--619, 2004.

\bibitem{Hartmann2003b}
M.~Hartmann, G.~Mahler, and O.~Hess.
\newblock Existence of {T}emperature on the {N}anoscale.
\newblock {\em Phys. Rev. Lett.}, (accepted), 2004.
\newblock quant-ph/0312214.

\bibitem{Hartmann2003}
M.~Hartmann, G.~Mahler, and O.~Hess.
\newblock Gaussian {Q}uantum {F}luctuations in {I}nteracting {M}any {P}article
  {S}ystems.
\newblock {\em math-ph/0312045}, 2004.
\newblock acc. for publ. in Lett. Math. Phys.

\bibitem{Hartmann2004a}
M.~Hartmann, G.~Mahler, and O.~Hess.
\newblock Local {V}ersus {G}lobal {T}hermal {S}tates: {C}orrelations and the
  {E}xistence of {L}ocal {T}emperatures.
\newblock {\em quant-ph/0404164}, 2004.

\bibitem{Hartmann2004b}
M.~Hartmann, G.~Mahler, and O.~Hess.
\newblock Spectral densities and partition functions of modular quantum systems
  as derived from a central limit theorem.
\newblock {\em cond-mat/0406100}, 2004.

\bibitem{Hill2001c}
T.L. Hill.
\newblock A {D}ifferent {A}pproach to {N}anothermodynamics.
\newblock {\em Nano Lett.}, 1(5):273, 2001.

\bibitem{Hill2001b}
T.L. Hill.
\newblock Extension of {N}anothermodynamics to {I}nclude a {O}ne-{D}imensional
  {S}urface {E}xcess.
\newblock {\em Nano Lett.}, 1(3):159, 2001.

\bibitem{Hill2001a}
T.L. Hill.
\newblock Perspective: {N}anothermodynamics.
\newblock {\em Nano Lett.}, 1(3):111, 2001.

\bibitem{Linnik1971}
I.A. Ibargimov and Y.V. Linnik.
\newblock {\em Independent and {S}tationary {S}equences of {R}andom
  {V}ariables}.
\newblock Wolters-Noordhoff, Groningen/Netherlands, 1971.

\bibitem{Jordan2003}
A.~N. Jordan and M.~B\"uttiker.
\newblock Entanglement {E}nergetics at {Z}ero {T}emperature.
\newblock {\em Phys. Rev. Lett.}, 92:247901, 2004.
\newblock cond-mat/0311647.

\bibitem{Kenzelmann2002}
M.~Kenzelmann, R.~Coldea, D.A. Tennant, D.~Visser, M.~Hofmann, R.~Smeibidl, and
  Z.~Tylczynski.
\newblock Order-to-disorder transition in the {XY}-like quantum magnet
  {C}s2{C}o{C}l4 induced by noncommuting applied fields.
\newblock {\em Phys. Rev. B}, 65:144432, 2002.

\bibitem{Kittel1983}
Ch. Kittel.
\newblock {\em Quantum {T}heory of {S}olids}.
\newblock Wiley, New York, 1963.

\bibitem{Kreuzer1984}
H.~J. Kreuzer.
\newblock {\em Nonequilibrium {T}hermodynamics and its statistical foundation}.
\newblock Clarandon Press, Oxford, 1981.

\bibitem{Kubo1985}
R.~Kubo, M.~Toda, and N.~Hashitsume.
\newblock {\em Statistical {P}hysics {II}: {N}onequilibrium {S}tatistical
  {M}echanics}.
\newblock Number~31 in Solid-State Sciences. Springer, Berlin, Heidelberg,
  New-York, 2. edition, 1991.

\bibitem{Lebowitz1969}
J.L. Lebowitz and E.H. Lieb.
\newblock Existence of {T}hermodynamics for {R}eal {M}atter with {C}oulomb
  {F}orces.
\newblock {\em Phys. Rev. Lett.}, 22:631, 1969.

\bibitem{Mahler1998}
G.~Mahler and V.A. Weberru\ss.
\newblock {\em Quantum Networks}.
\newblock Springer, Berlin, Heidelberg, 2. edition, 1998.

\bibitem{Michel2003}
M.~Michel, M.~Hartmann, J.~Gemmer, and G.~Mahler.
\newblock Fourier's {L}aw confirmed for a class of small quantum systems.
\newblock {\em Euro. Phys. J. B}, 34:325--330, 2003.

\bibitem{Nanda2003}
K.K. Nanda, A.~Maisels, F.E. Kruis, H.~Fissan, and S.~Stappert.
\newblock Higher {S}urface {E}nergy of {F}ree {N}anoparticles.
\newblock {\em Phys. Rev. Lett.}, 91:106102, 2003.

\bibitem{Nieuwenhuizen2002}
Th.~M. Nieuwenhuizen and A.~E. Allahverdyan.
\newblock Statistical {T}hermodynamics of {Q}uantum {B}rownian {M}otion:
  {C}onstruction of {P}erpetuum {M}obile of the {S}econd {K}ind.
\newblock {\em Phys. Rev. E}, page 036102, 2002.

\bibitem{Pothier1997}
H.~Pothier, S.~Gueron, N.O. Brige, D.~Esteve, and M.H. Devoret.
\newblock Energy {D}istribution {F}unction of {Q}uasiparticles in {M}esoscopic
  {W}ires.
\newblock {\em Phys. Rev. Lett.}, 79:3490, 1997.

\bibitem{Rajagopal2004}
A.K. Rajagopal, C.S. Pande, and {Sumiyoshi Abe}.
\newblock Nanothermodynamics - {A} generic approach to material properties at
  the nanoscale.
\newblock {\em cond-mat/0403738}, 2004.

\bibitem{Ruelle1969}
D.~Ruelle.
\newblock {\em Statistical {M}echanics}.
\newblock W.A. Benjamin Inc., New York, 1969.

\bibitem{Schmidt1998}
M.~Schmidt, R.~Kusche, B.~von Issendorf, and H.~Haberland.
\newblock Irregular variations in the melting point of size-selected atomic
  clusters.
\newblock {\em Nature}, 393:238, 1998.

\bibitem{Schwab2000}
K.~Schwab, E.A. Henriksen, J.M. Worlock, and M.L. Roukes.
\newblock Measurement of the quantum of thermal conductance.
\newblock {\em Nature}, 404:974--976, 2000.

\bibitem{Shinohara2003}
T.~Shinohara, T.~Sato, and T.~Taniyama.
\newblock Surface {F}erromagnetism of {P}d {F}ine {P}articles.
\newblock {\em Phys. Rev. Lett.}, 91:197201, 2003.

\bibitem{Tolman1967}
R.C. Tolman.
\newblock {\em The {P}rinciples of {S}tatistical {M}echanics}.
\newblock Oxford Univ. Press, London, 1967.

\bibitem{Varesi1998}
J.~Varesi and A.~Majumdar.
\newblock Scanning {J}oule expansion microscopy at nanometer scales.
\newblock {\em Appl. Phys. Lett.}, 72:37, 1998.

\bibitem{Wang2002}
X.~Wang.
\newblock Threshold temperature for pairwise and many-particle thermal
  entanglement in the isotropic heisenberg model.
\newblock {\em Phys. Rev. A}, 66:044305, 2002.

\bibitem{Williams1986}
C.C. Williams and H.K. Wickramasinghe.
\newblock Scanning thermal profiler.
\newblock {\em Appl. Phys. Lett.}, 49:1587, 1986.

\end{thebibliography}
